\documentclass[%
 preprint,
 amsmath,amssymb,
 aps,
 prx,
]{revtex4-1}

\usepackage{graphicx}
\usepackage{dcolumn}
\usepackage{bm}
\usepackage{color}
\usepackage{ulem}

\begin{document}

\title{A Parallel Beam Splitting Based on Gradient Metasurface:\\ Preparation and Fusion of Quantum Entanglement}

\author{Qi Liu$^{1,2}$}
\author{Xuan Liu$^3$}
\author{Yu Tian$^{1,2}$}
\author{Zhaohua Tian$^1$}
\author{Guixin Li$^4$}
\author{Xi-Feng Ren$^{5,6}$}
\author{Qihuang Gong$^{1,2,6,7,8}$}
\author{Ying Gu$^{1,2,6,7,8}$}
 \email{ygu@pku.edu.cn}

\affiliation{$^1$State Key Laboratory for Mesoscopic Physics, Department of Physics, Peking University, Beijing 100871, China\\
$^2$Frontiers Science Center for Nano-optoelectronics $\&$  Collaborative Innovation Center of Quantum Matter $\&$ Beijing Academy of Quantum Information Sciences, Peking University, Beijing 100871, China\\
$^3$ Institute of Laser Engineering, Faculty of Materials and Manufacturing, Beijing University of Technology, Beijing 100124 China\\
$^4$Department of Materials Science and Engineering, Southern University of Science and Technology, Shenzhen 518055, China\\
$^5$CAS Key Laboratory of Quantum Information, University of Science and Technology of China, Hefei 230026, China\\
$^6$Hefei National Laboratory, Hefei 230088, China\\
$^7$Collaborative Innovation Center of Extreme Optics, Shanxi University, Taiyuan, Shanxi 030006, China\\
$^8$Peking University Yangtze Delta Institute of Optoelectronics, Nantong 226010, China
}

\date{\today}
\begin{abstract}
Gradient metasurface, formed by a set of subwavelength unit cells with different phase modulation, is widely used in polarized beam splitting (BS) in the classical and quantum optics. 
Specifically, its phase gradient allows the path and polarization of multiple output lights to be locked by corresponding inputs.
Using this unique path-polarization locked property, we demonstrate that the single metasurface can function as sequentially linked beamsplitters, enabling the parallelization of a series of BS processes. Such a parallel BS metasurface provides a multi-beam interference capability for both classical and quantum light manipulation.
Taking this advantage, we first prepare path and polarization hybrid entangled states of two, three, and multi photons from unentangled photon sources. Then, the ability of parallel BS-facilitated entanglement is applied to demonstrate entanglement fusion among entangled photon pairs, which can greatly enlarge the entanglement dimension.
The principle of parallel BS through the metasurface opens up a versatile way to manipulate the quantum state at the micro/nano scale, which will have potential applications in on-chip quantum optics and quantum information processing.

\begin{description}
\item[Keywords]  Gradient metasurface;  Quantum beam splitting;  quantum entanglement
\end{description}

\end{abstract}


\maketitle

\section{Introduction}
Metasurface, a kind of artificial flat material composed of subwavelength elements~\cite{MSR1}, provides a nanoscale control to nearly all degrees of freedom of light~\cite{MSR2,MSR3,MSR4}. 
Its excellent light manipulation capability has made great success in polarization optics~\cite{MSPO2,MSPO3,MSPO}, nonlinear optics~\cite{MSNO}, structured light~\cite{MSSL1,MSSL2}, holograms~\cite{MSH1,MSH2}, metalens~\cite{Lens2,Lens}, and so on.
Recently, this unique property has been applied to integrated  quantum optics and quantum information processing~\cite{MQR}, such as 
high-dimensional and multiphoton entangled state generation~\cite{MQP1,MQP2,MQP3,MQP4}, quantum entanglement manipulation~\cite{MQM1,MQM2,MQM3,MQM4,MQM5,MQM6}, and quantum imaging~\cite{MQI1,MQI2,MQI3}.
Among them, the phase gradient metasurface has attracted much attention due to the ability to arbitrarily tailor the phase distribution of incident light~\cite{MQM5,MQM6,MQI1,MQI2,MQG1,MQG2,MQG3}.
Taking advantage of flexible phase gradient design, metasurfaces have been proposed to realize beam splitting (BS) functions with polarization-dependent ~\cite{MSBS1,MSBS2}, variable split ratio ~\cite{MSBS3,MSBS4}, and multi-channel~\cite{MSBS5,MSBS6} capabilities.
By utilizing the gradient metasurface BS,  quantum state reconstruction~\cite{MQBS1} and multichannel entanglement distribution/transformation~\cite{MQBS2} were demonstrated.
Also, owing to the polarization-dependent phase gradient, the path and polarization of outgoing light can be locked by the path and polarization of incident light~\cite{MSBS1,MSBS2}.
It implies that, in gradient metasurface,
if multiple beams from different paths are incident together,  the paths and polarizations of output beams can be locked by the corresponding input ones, and simultaneously they possess some common paths, i.e., these BS processes are parallel.
Although the parallel property of  BS in gradient metasurface has always existed, it has never been explored before. 
Once it is revealed, the concept of parallel BS will be used in various aspects of classical optics as well as in the quantum light manipulation. One of the purposes of our work is to setup the quantum frame of parallel BS of gradient metasurface.

Beam splitter is an indispensable tool for quantum state control in quantum optics and quantum information~\cite{QBS-R1,QBS-R2,QBS-R3}. 
A single beam splitter can enable the swapping~\cite{QBS-S1,QBS-S2,QBS-S3} and fusion~\cite{QBS-F1,QBS-F2} among two-party entanglement.
While a series of beam splitters together with phase controllers can construct a multiport quantum network for performing arbitrary linear unitary operations on a quantum state~\cite{QBS-N1,QBS-N2,QBS-N3,QBS-N4}. 
Through several beam splitters, the Greenberger-Horne-Zeilinger (GHZ) state with 12 photons~\cite{QBS-P1} and cluster state with more than 20000 modes~\cite{QBS-P2,QBS-P3} have been fabricated.
However, the use of too many beam splitters in quantum operations may bring some problems such as loss and crosstalk, which limits the integrability of these devices.
Hence, using the parallel feature of BS mentioned above, the metasurface will greatly reduce the number of devices.
Another purpose of our work is to employ gradient metasurface to fabricate quantum entanglement through nanoscale parallel BS, which will meet the demand for on-chip quantum optics and quantum information processing.

In this paper, we first propose the principle of the parallel BS  of a Pancharatnam-Berry(PB)-phase gradient metasurface. 
We model the phase gradient metasurface as a series of parallel beam splitters, each with the same splitting ratio, in both classical and quantum regions. These splitters are linked by sharing one common output path between two adjacent splitters, while the two splitters contribute a pair of orthogonal polarization modes on their common path. 
Secondly, we show that multiphoton entangled states can be realized based on the quantum interference induced by parallel BS. With a suitable selection of input paths from parallel BS processes, a genuine $N$-photon entangled state can be generated from unentangled input photons with a certain successful probability.
Subsequently, the ability to generate entanglement is further applied to the fusion of entanglement among multiple parties. Specifically, $N$ pairs of Bell entangled states are merged into a genuine $2N$-photon entangled state after the quantum interference on the metasurface, even though some of these photons never interact with each other.
Our approach provides a compact, simple, and efficient way to manipulate and generate multi-photon states. This method also enables multi-functional quantum state manipulations without additional modulation to metasurface devices. These findings may pave the way to future on-chip parallel quantum information processing.

The paper is organized as follows. In Sec. II, we setup the physical model of the parallel BS metasurface for classical and quantum light.
 Then, in Section III,  using the parallel BS, quantum state manipulations are discussed, including the preparation of entangled state and entanglement fusion. Finally, we summarized our work in Sec. IV.

\section{The parallel BS model of metasurface}
\begin{figure}[hbtp]
  \centering
  \includegraphics[width=0.5\textwidth]{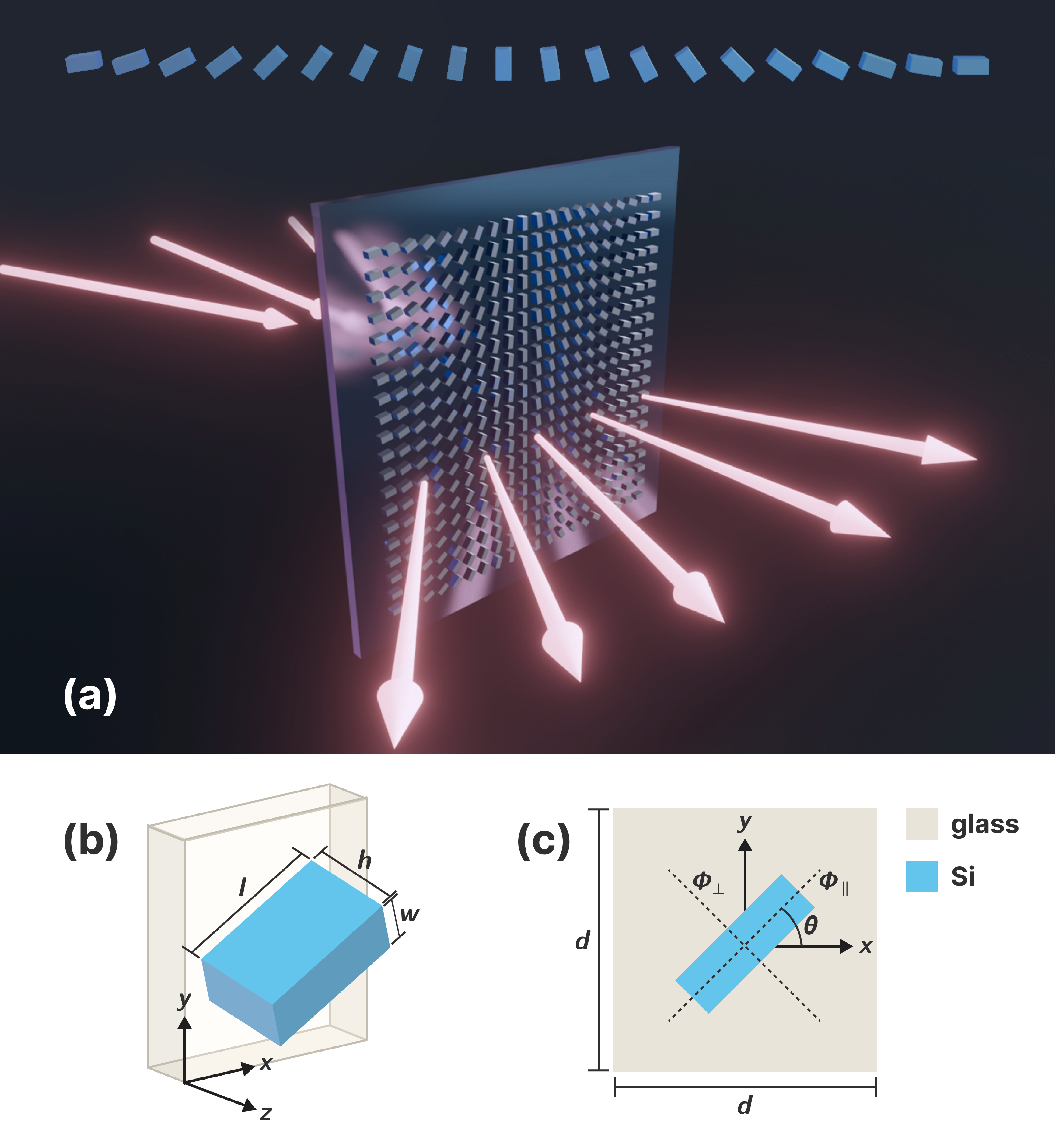}\\
  \caption{(a) The schematic of gradient metasurface with the function of parallel BS. Through these BSs, the polarizations and paths of multiple output beams are locked by those corresponding input beams.
  (b) The unit cell is composed of a silicon nanofin (blue) and a glass substrate (gray). The size of the nanofin ($w,l,h$) determines the local birefringent response of the metasurface. (c) The top view of a unit cell. Each nanofin is rotated around $z$-axis by an angle $\theta$, which varies with the position. The neighboring nanofins are separated by $d$.
}\label{fig1}
\end{figure}
Recent advances show that metasurfaces can provide unique BS capability for classical~\cite{MSBS1,MSBS2,MSBS3,MSBS4,MSBS5,MSBS6} and quantum~\cite{MQM2,MQM3,MQBS1,MQBS2} applications.
 In classical region, one incident light is split into two or more paths for realizing multiplexing and classical interference~\cite{BS-book}. 
 While in quantum region, after passing through the metasurface,  quantum state can be constructed through the interference effect~\cite{QBS-book}. 
 However, most of these advances are focused on one-to-multi BS situations, the possibility of manipulating quantum states with multi-to-multi BS remains unclear.
In this section, we mainly propose the principles of parallel BS on a gradient metasurface for classical light and quantum light.


\subsection{BS rule for single-input light}
Consider a PB-phase metasurface constructed by a periodic array of dielectric nanofins with different oriented angle $\theta(x,y)$ [Fig.~\ref{fig1} (a)]. The geometry of the unit cell is shown in Fig.~\ref{fig1} (b), where a high-index dielectric (such as silicon) nanofin with the length $l$, width $w$, and height $h$ sits on a glass substrate. The neighboring nanofins are separated by $d$. Here $l,w,d$ are great less than the wavelength $\lambda$.
Due to the rectangle geometry of nanofin in $xy$-plane, the nanofin can act as a birefringent element when the light propagates along $z$-axis through the unit cell.
As shown in Fig.~\ref{fig1} (c), a nanofin oriented at an angle $\theta$ will impose phase shifts $\phi_{\parallel}, \phi_{\perp}$ when a linearly polarized light is added~\cite{MSR2}. 
By assuming that the dielectric is lossless and local field coupling among nearby unit cells is negligible, the linearly birefringent response of the unit cell in Fig.~\ref{fig1} (c) can be described by a local Jones matrix~\cite{MSBS1,MSBS2}
\begin{equation}\label{Jones}
\begin{split}
J(\theta)&=R(\theta)
\begin{bmatrix}
	e^{i\phi_{\parallel}} & 0\\
	0                     & e^{i\phi_{\perp}}
\end{bmatrix}
R(-\theta)\\
&=e^{i\phi_D}\cos \Delta J_0+ie^{i\phi_D}\sin \Delta (e^{+2i\theta}J_{+}+e^{-2i\theta}J_{-}),
\end{split}
\end{equation}
where $R(\theta)$ is a $2\times2$ rotation matrix, $J_0$ is an identity matrix, $J_{+/-}$ is a Jones matrix which converts left-circularly polarized (LCP) / right-circularly polarized (RCP) light into RCP / LCP light, $\phi_D=(\phi_{\parallel}+\phi_{\perp})/2$, $\Delta=(\phi_{\parallel}-\phi_{\perp})/2$.
 If we define the Jones vectors as $ |L\rangle =[1,i]^T/\sqrt{2}$, $|R\rangle =[1,-i]^T/\sqrt{2}$ ($T$ denotes transpose), then $J_0=|L\rangle\langle L|+|R\rangle\langle R|$, $J_{+}=|R\rangle\langle L|$, $J_{-}=|L\rangle\langle R|$. Equation~(1) indicates that the unit cell can control the amplitude and phase simultaneously for circularly polarized light. The phase retardation difference $\Delta$ can control the proportion of light that flips handedness of circular polarization state, while the oriented angle $\theta$ induces an additional phase $+2\theta/-2\theta$ of flipped LCP/RCP incident light, known as PB phase or geometry phase.

The BS capability of metasurface can be realized by employing PB-phase gradient~\cite{MSBS1,MSBS2}. 
Without loss of generality, we assume a phase gradient is along $x$-axis, so $\theta(x,y)=k_{_G}x/2$ with $k_{_G}=2\pi/(Md)$, where $M$ is the number of unit cell in one period (typically, $M>5$). In our analyses, $M$ is large enough to guarantee that all splitting lights satisfy paraxial approximation. Then, the Jones matrix  varying with coordinate ($x,y$) is
\begin{equation}\label{Jones2}
J(x,y)=\cos \Delta J_0+i\sin \Delta (e^{+ik_{_G}x}J_{+}+e^{-ik_{_G}x}J_{-}).
\end{equation}
Note that the global phase factor $\exp(i\phi_D)$ has been omitted in Eq.~(\ref{Jones2}). The wavefronts before and after the light passing through the metasurface are connected by $|\mathbf{E}_{out}(x,y)\rangle=J(x,y)|\mathbf{E}_{in}(x,y)\rangle$. This relation suggests that one input light is divided into at most three parts, since each part of light feels a different phase gradient ($-k_{_G},0,+k_{_G}$).

\begin{figure*}[htbp]
  \centering
  \includegraphics[width=0.95\textwidth]{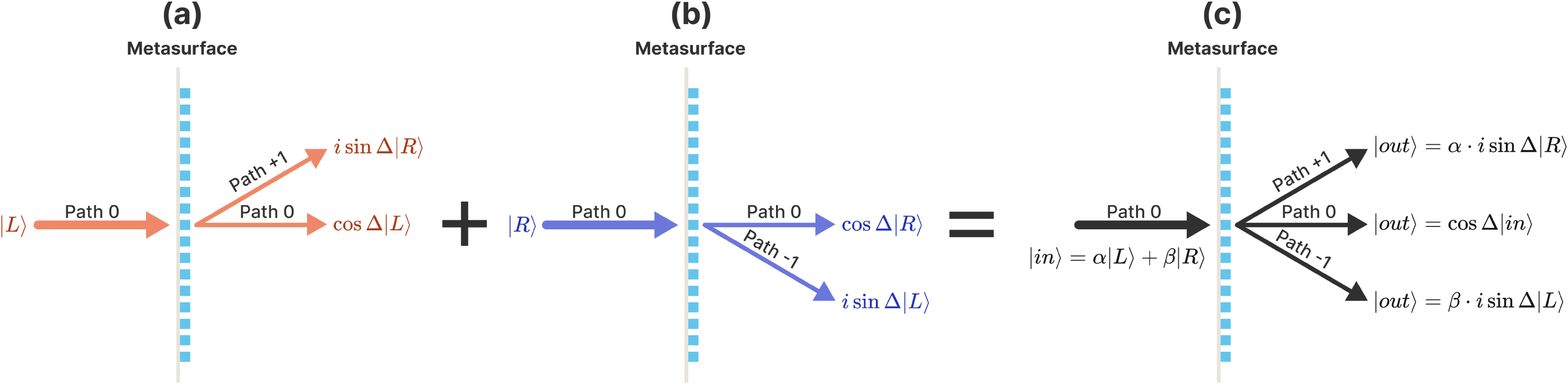}\\
  \caption{Beam splitting rules on a PB-phase gradient metasurface. BS processes for (a) left-circularly and (b) right-circularly polarized lights. Some part of the light will not change its propagating direction (path 0) and preserve its polarization, while the remaining part will be deflected into paths $+1$  or $-1$ with its polarization flipped into the opposite one. (c) BS process for an arbitrarily polarized light, which can be regarded as the superposition of (a) and (b). In (c),  the light at most can be split into three paths; while in (a) and (b), at most two paths. There is a phase difference $\pi/2$ between two split paths in (a) and (b).
  }\label{fig2}
\end{figure*}

With such a phase gradient, the circularly polarized light can be split into two paths. As shown in Fig.~\ref{fig2} (a), for a normal incident LCP light (thick red arrow along path 0), after passing through the metasurface, one part of the light converts into RCP with phase gradient $k_{_G}$, so it deflected into a new path +1. 
Another part keeping its polarization handedness propagates along the same path as the incident light. 
The split ratio of amplitude is $\tan \Delta$. 
Similarly, for an incident RCP light, it will also be split into two beams [Fig.~\ref{fig2} (b)]. But the part converted into LCP feels an opposite phase gradient ($-k_{_G}$), thus this part deflects along an opposite direction with the same angle. 

For an incident light with polarization $|{\rm in}\rangle_0=\alpha|L_{in}\rangle_0+\beta|R_{in}\rangle_0$, the LCP and RCP components obey the above splitting rule. It will be generally split into three beams with different polarizations as illustrated in Fig.~\ref{fig2} (c). Then, the BS  of the metasurface can be expressed as
\begin{equation}\label{BS1}
\begin{bmatrix}
|{\rm out}\rangle_{+1}\\
|{\rm out}\rangle_{0}~\\
|{\rm out}\rangle_{-1}
\end{bmatrix}
=
\begin{bmatrix}
\tilde{J}_{+}\\
\tilde{J}_0\\
\tilde{J}_{-}
\end{bmatrix}
|{\rm in}\rangle_{0},
\end{equation}
with $\tilde{J}_0=\cos\Delta\cdot J_{0}$, $\tilde{J}_{\pm}=i\sin\Delta\cdot J_{\pm}$. Here, $|{\rm out(in)}\rangle_{j}$ is the Jones vector of the output (input) light propagating along the path $j$. In path $j$,  $\mathbf{E}_j=|{\rm out(in)}\rangle_{j}\cdot\exp(i\mathbf{k}_j\cdot\mathbf{r})$, where $\mathbf{r}=[x,y,z]^T$, $\mathbf{k}_j=[jk_{_G},0,\sqrt{k_0^2-(jk_{_G})^2}]^T(j=0,\pm1,...)$.


\subsection{Parallel BS for classical light}
\begin{figure*}[htbp]
  \centering
  \includegraphics[width=0.95\textwidth]{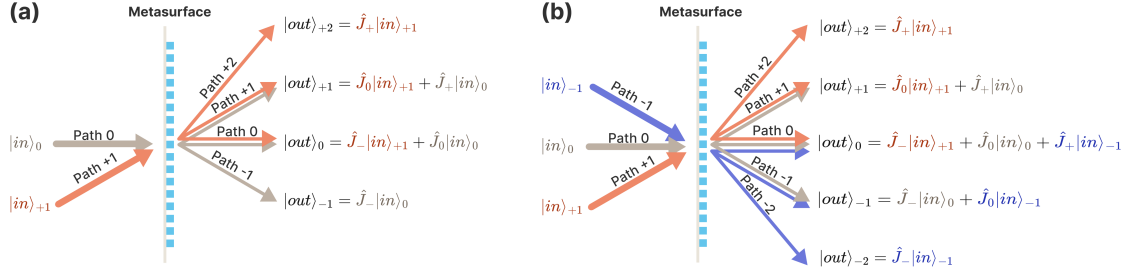}\\
  \caption{Basic working principle of parallel BS in PB-phase gradient metasurface. 
  (a) Parallel BS for two incident beams from paths 0 and $+1$. Two BS processes yield 4 output paths ($\pm1,0,+2$) with 2 common output paths ($0,+1$). 
  (b) Parallel BS for three incident beams from paths 0 and  $\pm1$. Three BS processes yield 5 output paths ($0, \pm1,\pm2$) with 3 common output paths ($0,\pm1$). 
  }\label{fig3}
\end{figure*}

We first consider classical parallel BS model. The path modes are plane waves with wavevectors $\mathbf{k}_j=[jk_{_G},0,\sqrt{k_0^2-(jk_{_G})^2}]^T(j=0,\pm1,\pm2...)$, whose polarization states can be described by an arbitrary Jones vector. So two adjacent path modes differ by a metasurface phase gradient $k_{_G}$. 
When two beams are incident from paths 0 and $+1$, their parallel BS process is shown in Fig.~\ref{fig3} (a).
According to Eq.~(\ref{Jones2}), each beam is split into three paths, then two sets of output paths yield
4  beams (paths $-1,0,+1,+2$). Two of them (paths $0,+1$) are a coherent mixing of incident beams. Such a parallel BS process is described by
\begin{equation}\label{BS2}
\begin{bmatrix}
|{\rm out}\rangle_{+2}\\
|{\rm out}\rangle_{+1}\\
|{\rm out}\rangle_{0}~\\
|{\rm out}\rangle_{-1}
\end{bmatrix}
=
\begin{bmatrix}
\tilde{J}_{+}&0\\
\tilde{J}_{0}&\tilde{J}_{+}\\
\tilde{J}_{-}&\tilde{J}_{0}\\
0&\tilde{J}_{-}
\end{bmatrix}
\begin{bmatrix}
|{\rm in}\rangle_{+1}\\
|{\rm in}\rangle_{0}~
\end{bmatrix}.
\end{equation}
So in this two-input parallel BS process, output polarization states in paths 0 and $+1$ are a coherent superposition of two input states transformed by different Jones matrixes.

It is easy to extend two parallel BS to three parallel BS processes. As shown in Fig.~\ref{fig3} (b), when three beams are incident from paths $0,\pm1$, similarly,  three sets of output paths overlapped with each other, producing five output paths ($0, \pm1, \pm2$). Two of these paths $\pm1$ are coherently mixed by two incident beams, while path 0 is mixed by all three beams. The parallel BS process can be described as
\begin{equation}\label{BS3}
\small
\begin{bmatrix}
|{\rm out}\rangle_{+2}\\
|{\rm out}\rangle_{+1}\\
|{\rm out}\rangle_{0}~\\
|{\rm out}\rangle_{-1}\\
|{\rm out}\rangle_{-2}
\end{bmatrix}
=
\begin{bmatrix}
\tilde{J}_{+}&0&0\\
\tilde{J}_{0}&\tilde{J}_{+}&0\\
\tilde{J}_{-}&\tilde{J}_{0}&\tilde{J}_{+}\\
0&\tilde{J}_{-}&\tilde{J}_{0}\\
0&0&\tilde{J}_{-}
\end{bmatrix}
\begin{bmatrix}
|{\rm in}\rangle_{+1}\\
|{\rm in}\rangle_{0}~\\
|{\rm in}\rangle_{-1}
\end{bmatrix}.
\end{equation}
Using polarization bases $|L\rangle$ and $|R\rangle$,  if only $|R_{\rm in}\rangle_{+1}$ and $|L_{\rm in}\rangle_{-1}$ exist for paths $\pm1$ input, the corresponding $|L_{out}\rangle$, $|R_{out}\rangle$ are written as 
\begin{equation}\label{BS3-1}
\small
\begin{bmatrix}
|R_{\rm out}\rangle_{+1}\\
|L_{\rm out}\rangle_{0}\\
|R_{\rm out}\rangle_{0}\\
|L_{\rm out}\rangle_{-1}
\end{bmatrix}
=
\begin{bmatrix}
\cos\Delta&i\sin\Delta&0&0\\
i\sin\Delta&\cos\Delta&0&0\\
0&0&\cos\Delta&i\sin\Delta\\
0&0&i\sin\Delta&\cos\Delta
\end{bmatrix}
\begin{bmatrix}
|R_{\rm in}\rangle_{+1}\\
|L_{\rm in}\rangle_{0}\\
|R_{\rm in}\rangle_{0}\\
|L_{\rm in}\rangle_{-1}
\end{bmatrix},
\end{equation}
where path 0 shows the feature of parallel BS. Eq. (6) can be directly generalized to the quantum situation.
From above examples, we can see that the parallel BS provides a more flexible way to control the output beams, including their polarization, phase and intensity.
This is obviously superior to single-input BS on metasurface, where the polarization of output beams is only transformed from one BS process by a certain Jones matrix~\cite{MSPO,MSBS1,MSBS2,MSBS6}. 

\begin{figure}[htb]
  \centering
  \includegraphics[width=0.45\textwidth]{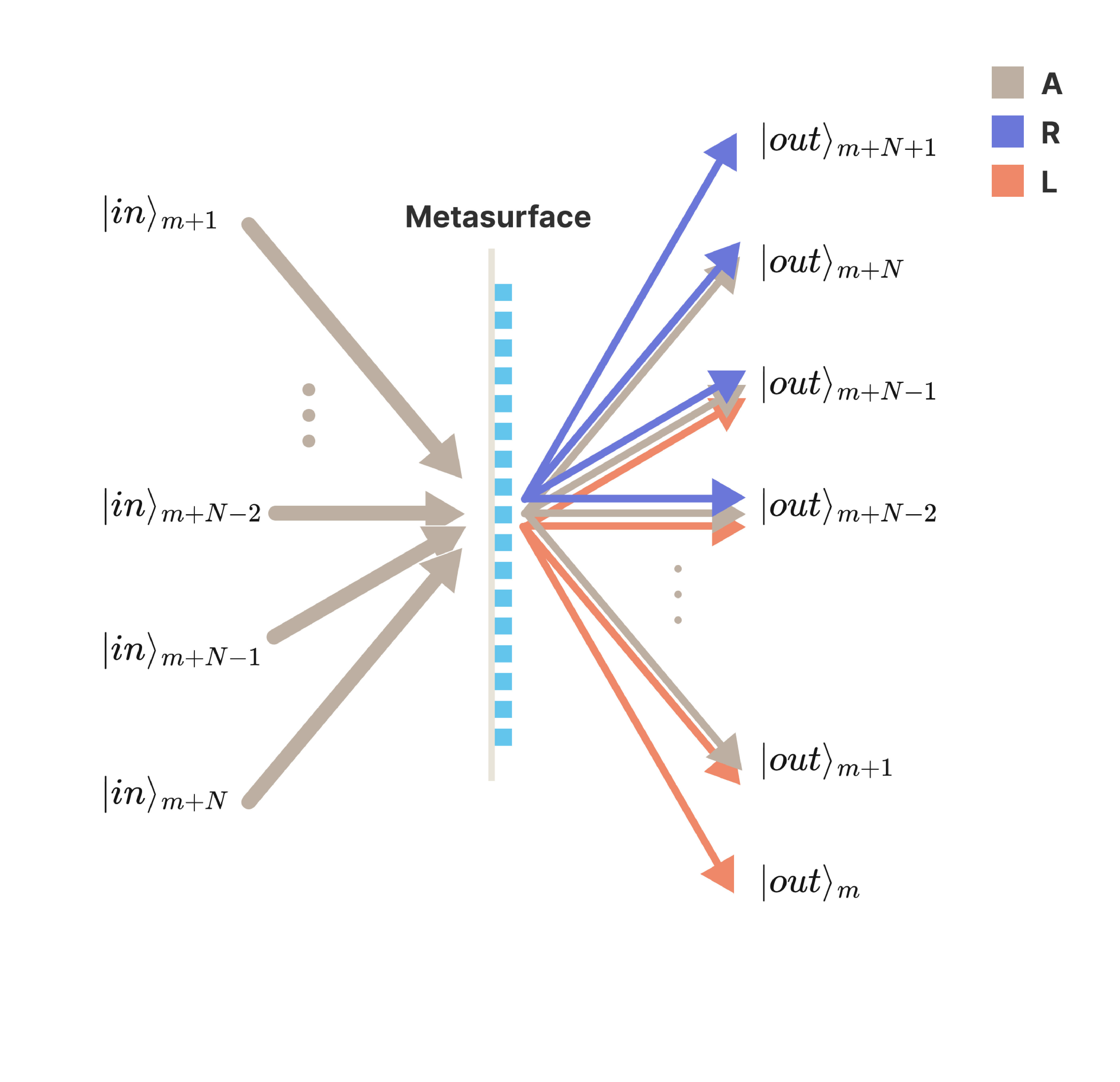}\\
  \caption{A general scheme for parallel BS metasurface with N ($>3$) incident path. The incident beams with arbitrary polarization can be different from each other.
  The output beam in one path may be a mixing of three/two/one (corresponding to the number of arrows in the path) incident beams.
}\label{fig4}
\end{figure}

Subsequently, the parallel BS rule of gradient metasurface can be extended to the cases of multiple incident lights. 
As shown in Fig.~\ref{fig4}, multiple beams are incident from $N$ paths,  yielding $N+2$ output paths. 
The output polarizations (LCP/A/RCP) and paths [$(j-1)/j/(j+1)$] are locked by the incident light (path $j$ of polarization state A) for each parallel BS process. 
Furthermore, the phase gradient of metasurface ensures that the output paths are overlapped as incident beams are from adjacent paths. 
Hence, for  a general gradient metasurface, its parallel BS rule is described by
\begin{widetext}
\begin{equation}\label{BS4}
\small
\begin{bmatrix}
|{\rm out}\rangle_{m+N+1}\\
|{\rm out}\rangle_{m+N}~~~\\
|{\rm out}\rangle_{m+N-1}\\
~\vdots~\\
|{\rm out}\rangle_{m+2}\\
|{\rm out}\rangle_{m+1}\\
|{\rm out}\rangle_{m}~~~
\end{bmatrix}_{(N+2)\times1}
=
\begin{bmatrix}
\tilde{J}_{+}&~&~&~&~\\
\tilde{J}_{0}&\tilde{J}_{+}&~&~&~\\
\tilde{J}_{-}&\tilde{J}_{0}&\ddots&~&~\\
~&\tilde{J}_{-}&\ddots&\tilde{J}_{+}&~\\
~&~&\ddots&\tilde{J}_{0}&\tilde{J}_{+}\\
~&~&~&\tilde{J}_{-}&\tilde{J}_{0}\\
~&~&~&~&\tilde{J}_{-}
\end{bmatrix}_{(N+2)\times N}
\begin{bmatrix}
|{\rm in}\rangle_{m+N}~~~\\
|{\rm in}\rangle_{m+N-1}\\
~\vdots~\\
|{\rm in}\rangle_{m+2}\\
|{\rm in}\rangle_{m+1}
\end{bmatrix}_{N\times1},
\end{equation}
\end{widetext}
and if using polarization bases $|L\rangle$ and $|R\rangle$, it has the form
\begin{widetext}
\begin{equation}\label{BS4-1}
\tiny
\begin{bmatrix}
|R_{\rm out}\rangle_{m+N}~~~\\
|L_{\rm out}\rangle_{m+N-1}\\
|R_{\rm out}\rangle_{m+N-1}\\
|L_{\rm out}\rangle_{m+N-2}\\
|R_{\rm out}\rangle_{m+N-2}\\
~\vdots~\\
|L_{\rm out}\rangle_{m+2}\\
|R_{\rm out}\rangle_{m+2}\\
|L_{\rm out}\rangle_{m+1}
\end{bmatrix}
=
\begin{bmatrix}
\cos\Delta&i\sin\Delta&~&~&~&~&~&~&~\\
i\sin\Delta&\cos\Delta&~&~&~&~&~&~&~\\
~&~&\cos\Delta&i\sin\Delta&~&~&~&~&~\\
~&~&i\sin\Delta&\cos\Delta&~&~&~&~&~\\
~&~&~&~&\cos\Delta&i\sin\Delta&~&~&~\\
~&~&~&~&~&\ddots&~&~&~\\
~&~&~&~&~&i\sin\Delta&\cos\Delta&~&~\\
~&~&~&~&~&~&~&\cos\Delta&i\sin\Delta\\
~&~&~&~&~&~&~&i\sin\Delta&\cos\Delta
\end{bmatrix}
\begin{bmatrix}
|R_{\rm in}\rangle_{m+N}~~~\\
|L_{\rm in}\rangle_{m+N-1}\\
|R_{\rm in}\rangle_{m+N-1}\\
|L_{\rm in}\rangle_{m+N-2}\\
|R_{\rm in}\rangle_{m+N-2}\\
~\vdots~\\
|L_{\rm in}\rangle_{m+2}\\
|R_{\rm in}\rangle_{m+2}\\
|L_{\rm in}\rangle_{m+1}
\end{bmatrix}.
\end{equation}
\end{widetext}
By sharing the same path of both LCP and RCP, a parallel BS including $N-2$ splittings is setup. 
Here, the maximum number of the input/output light beam is mainly constrained by paraxial approximation. When the oblique incident angle is too large to obey the approximation, Eq.~(\ref{Jones}) becomes inaccurate thus the derived parallel BS relations~(\ref{BS4} and 8) are not available anymore. By decreasing the phase gradient $k_{_G}$, the splitting angle is reduced thus more light beams can incident. 

Previous studies only focus on single-input BS and one kind of polarization manipulation~\cite{MSBS5,MSBS6,MQBS1,MQBS2}, which usually requires complex metasurface design thus increasing the integrated difficulty. 
 While in a parallel BS, owing to the multiple light interferences on metasurface, the parallel BS further improves the degree of freedom for coherent control of polarization state.
 Hence the implementation of parallel BS can relax the complexity of metasurface design, maintaining flexibility for polarization control at the same time.
Nevertheless, the metasurface we analyzed here can only split one incident light beam into at most three paths, which limits the number of light beams that can be coherently superposed in one path. 
 Utilizing spatial multiplexing~\cite{MQBS1} or Jones matrix Fourier optics~\cite{MSPO,MSBS6} strategies, the BS paths for single input can be greatly increased. Thus, a richer feature in parallel BS may be available.
This multi-beam interference property of parallel BS can also be extended to the quantum region, which will be explored in the next Subsection.

\subsection{Parallel BS for quantum light}
\begin{figure*}[htbp]
  \centering
  \includegraphics[width=0.99\textwidth]{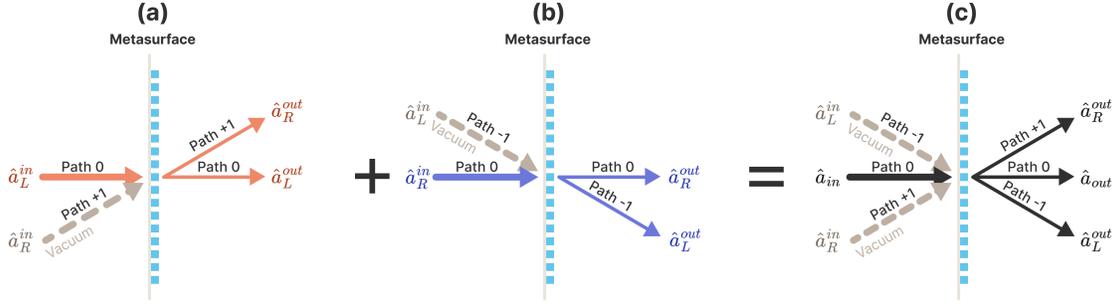}\\
  \caption{BS rule of metasurface for single incident quantum light. 
  Splitting process for (a)  left-circularly,  (b)  left-circularly, or  (c) arbitrary polarized light incident from path 0. 
  Input vacuum paths $\pm1$ are supplied to guarantee that input and output creation and annihilation operators satisfy the commutation relation. 
      (c)  process can be regarded as a proper superposition of (a) and (b) processes. 
    }\label{fig5}
\end{figure*}
Before the discussion of the parallel BS in quantum region, let us appoint the quantum description of the input and output lights. 
The LCP and RCP mode propagates along the path $j$ are denoted by annihilation operator $\hat{a}_{L(R)}^{in}(j)$ or $\hat{a}_{L(R)}^{out}(j)$, which satisfy the Bosonic commutation relations $[\hat{a}_{A}^{in(out)}(j), \hat{a}_{B}^{in(out)}(k)]=0$, $[\hat{a}_{A}^{in(out)}(j), \hat{a}_{B}^{\dag in(out)}(k)]=\delta_{j,k}\delta_{A,B}$ ($A,B\in\{L,R\}$). 
For an arbitrary polarization in the path $j$, its annihilation operator $\hat{a}_{in(out)}(j)$ is a superposition of LCP and RCP mode: $\hat{a}_{in(out)}(j)=\alpha_j\hat{a}_{L}^{in(out)}(j)+\beta_j\hat{a}_{R}^{in(out)}(j)$ with $|\alpha_j|^2+|\beta_j|^2=1$, which also satisfies the Bosonic commutation relations. 
Even if the mode is not occupied, i.e., quantum vacuum, the vacuum fluctuation will influent the quantum BS result~\cite{QBS-book}.

We first analyze the quantum parallel BS with only one input light. 
As illustrated in Fig.~\ref{fig5} (a) or (b), only LCP or RCP mode of path 0, denoted by $\hat{a}_{L}^{in}(0)$ or $\hat{a}_{R}^{in}(0)$, is occupied by photons in the input port. 
Such a quantum input yields two output modes for path $0,+1$ or $(0,-1)$ [thin arrows in Fig.~\ref{fig5} (a) or (b)], corresponding to annihilation operators $\hat{a}_{L}^{out}(0), \hat{a}_{R}^{out}(+1)$ for LCP input [or $\hat{a}_{R}^{out}(0), \hat{a}_{L}^{out}(-1)$ for RCP], but the vacuum mode $\hat{a}_{R }^{in}(+1)$ or $\hat{a}_{L }^{in}(-1)$  also takes part in the BS process.
The BS process in Fig.~\ref{fig5} (a) or (b) can be described by an interaction Hamiltonian as
\begin{equation}\label{H1}
\hat{H}_j=-g\hbar\left[\hat{a}_{L}^{\dag in}(j)\hat{a}_{R}^{in}(j+1)+\hat{a}_{L}^{in}(j)\hat{a}_{R}^{\dag in}(j+1)\right],
\end{equation}
where $g$ is the effective coupling strength determined by split ratio of metasurface, $j=0$ for LCP and $j=-1$ for RCP.
As shown in Fig.~\ref{fig5} (c), if in path 0 the input is a general polarization, i.e., $\hat{a}_{in}(0)=\alpha_0\hat{a}_{L}^{in}(0)+\beta_0\hat{a}_{R}^{in}(0)$.
Then its effective interaction Hamiltonian is $\hat{H}_{eff}=\hat{H}_0+\hat{H}_{-1}$, which involves three input paths $0,\pm1$. 
In this case, even though only one path is occupied by photons, three input paths should be taken into account together for the requirement of the parallel BS.

Once the effective Hamiltonian has been setup, 
the time evolution operator of this parallel BS reads $\hat{S}(t)=\exp(-i\hat{H}_{eff}t/\hbar)$, where $t$ is effective interaction time. 
In Heisenberg picture, the evolution from input mode to output modes is given by $\hat{a}_{out}(j)=\hat{S}^\dag(t)\hat{a}_{in}(j)\hat{S}(t)$~\cite{QBS-book}, then we obtain the input-output transformation for circularly polarized modes
\begin{subequations}\label{QBS1}
\small
\begin{align}
&\hat{a}_{R}^{out}(j)=\cos\Delta\cdot\hat{a}_{R}^{in}(j)+i\sin\Delta\cdot\hat{a}_{L}^{in}(j-1),j=0,+1\\
&\hat{a}_{L}^{out}(j)=\cos\Delta\cdot\hat{a}_{L}^{in}(j)+i\sin\Delta\cdot\hat{a}_{R}^{in}(j+1),j=0,-1
\end{align}
\end{subequations}
with $\Delta=gt$ only for the form, which in fact represents the split radio provided by the metasurface. 
Then, when $\hat{a}_{in}(0)=\alpha_0\hat{a}_{L}^{in}(0)+\beta_0\hat{a}_{R}^{in}(0)$, $\hat{a}_{out}(0)$ in path 0 reads 
\begin{eqnarray}\label{QBS2}
&\hat{a}_{out}(0)=\cos\Delta\cdot\hat{a}_{in}(0)+i\alpha_0\sin\Delta\cdot\hat{a}_{R}^{in}(+1)
+i\beta_0\sin\Delta\cdot\hat{a}_{L}^{in}(-1).
\end{eqnarray}
Formally, the BS process of Fig. 5(c) can be expressed as 
\begin{equation}\label{QBS3}
\small
\begin{bmatrix}
\hat{a}_{R}^{out}(+1)\\
\hat{a}_{out}(0)\\
\hat{a}_{L}^{out}(-1)
\end{bmatrix}
=
\begin{bmatrix}
\hat{J}_{0} & \hat{J}_{+} & 0\\
\alpha_0\hat{J}_{-} & \hat{J}_{0} & \beta_0\hat{J}_{+}\\
0 & \hat{J}_{-} & \hat{J}_{0}
\end{bmatrix}
\begin{bmatrix}
\hat{a}_{R}^{in}(+1)\\
\hat{a}_{in}(0)\\
\hat{a}_{L}^{in}(-1)
\end{bmatrix},
\end{equation}
where $\hat{J}_0=\cos \Delta, \hat{J}_{\pm}=i\sin\Delta\cdot\hat{P}_{\pm}$. Here, $\hat{P}_{\pm}$ is projection “operator”, which project any annihilation $\hat{a}_{in}(j)$ into its LCP or RCP “component”, {\rm i.e.}, $\hat{P}_{+}\hat{a}_{in}(j)= \hat{a}_{L}^{in}(j)$, $\hat{P}_{-}\hat{a}_{in}(j)= \hat{a}_{R}^{in}(j)$. Note that Eq~(\ref{QBS3}) is not a standard notation for quantum BS transformation, because it merges two orthogonal polarization modes [$\hat{a}_{L}^{in}(0)$ and $\hat{a}_{R}^{in}(0)$] into one arbitrary polarization mode [$\hat{a}_{in}(0)$], at a cost of using additional projection operators $\hat{P}_{\pm}$. But this notation clearly reveals that the quantum BS for an arbitrary polarization mode incident contains two parallel BS processes. 
From Fig. 5, one can see that, (c) is a complete quantum BS process, by sharing the common path 0, which includes two splitting precesses (a) and (b). So it is a most simple parallel quantum BS.
By using polarization bases $|L\rangle$ and $|R\rangle$, 
a more natural way to describe above parallel BS process is 
\begin{equation}\label{QBS4}
\small
\begin{bmatrix}
\hat{a}_{R}^{out}(+1)\\
\hat{a}_{L}^{out}(0)\\
\hat{a}_{R}^{out}(0)\\
\hat{a}_{L}^{out}(-1)
\end{bmatrix}
=
\begin{bmatrix}
\cos \Delta & i\sin\Delta & 0 & 0\\
i\sin\Delta & \cos \Delta & 0 & 0\\
0 & 0 & \cos \Delta & i\sin\Delta\\
0 & 0 & i\sin\Delta & \cos \Delta
\end{bmatrix}
\begin{bmatrix}
\hat{a}_{R}^{in}(+1)\\
\hat{a}_{L}^{in}(0)\\
\hat{a}_{R}^{in}(0)\\
\hat{a}_{L}^{in}(-1)
\end{bmatrix}.
\end{equation}
In this way, two non-zero block $2\times2$ matrices show the two parallel BS processes, which also indicates each process involves two circularly polarized modes.  One can easily check that both transformations in (\ref{QBS3}) and (\ref{QBS4}) maintain the Bosonic commutation relations.

\begin{figure}[htbp]
  \centering
  \includegraphics[width=0.6\textwidth]{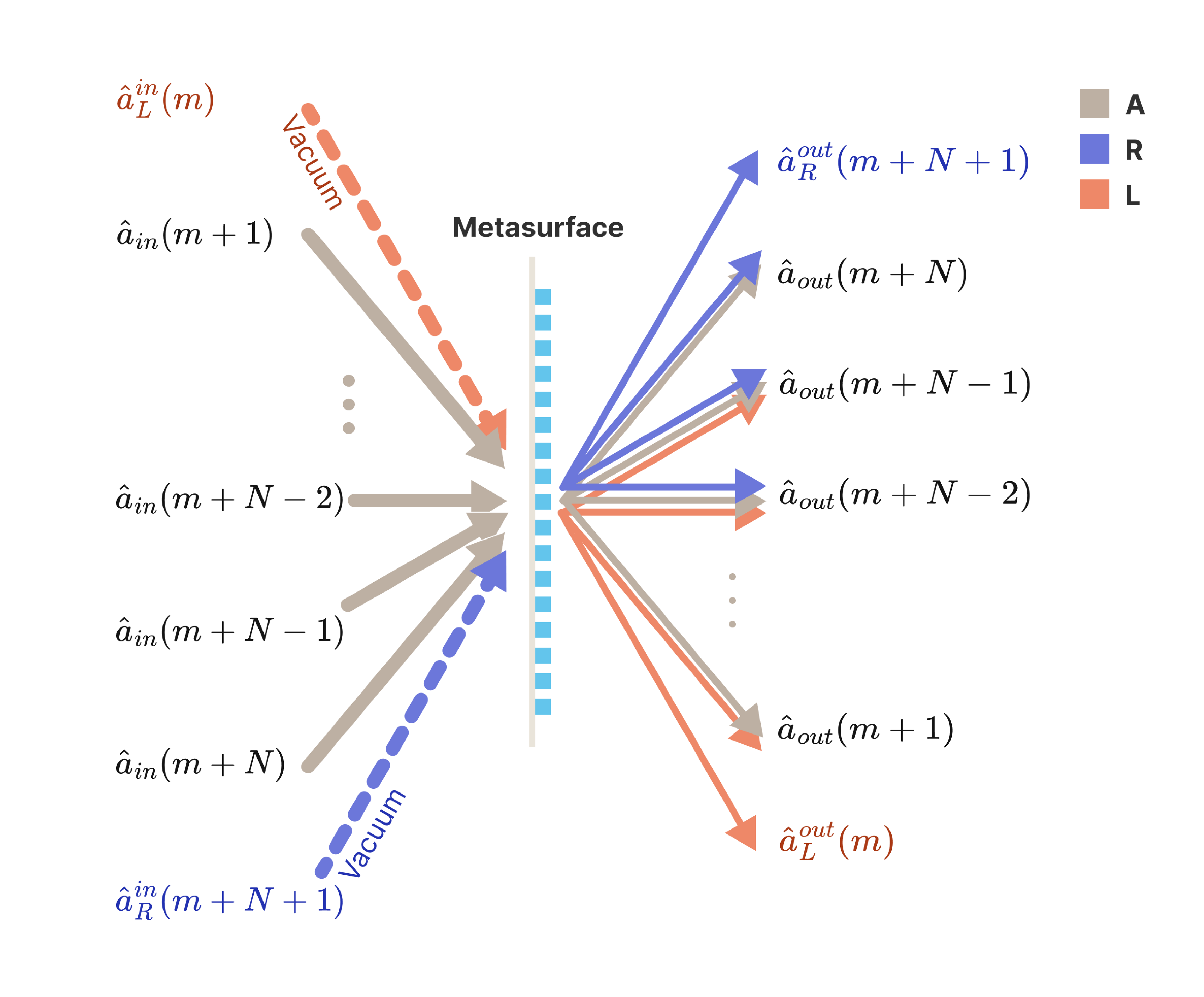}\\
  \caption{A parallel BS of gradient metasurface for multiple quantum lights.  
  Lights are incident from $N$ different paths, each of whose polarizations is the superposition of LCP and RCP. 
  To satisfy the commutation relation, the vacuum inputs from path $m$ (LCP) and path $m+N+1$ (RCP) are added.
}\label{fig6}
\end{figure}

Next, we consider the quantum parallel BS process of multiple incident lights. 
As shown in Fig.~\ref{fig6}, there are $N$ input path modes with the arbitrary polarization, indexed  from $(m+1)$ to $(m+N)$.
As a result, $(N+2)$ output path modes are occupied, indexed from $m$ to $(m+N+1)$.
We notice that modes $\hat{a}_{L}^{out}(m), \hat{a}_{R}^{out}(m+1)$ may be contributed by input mode $\hat{a}_{L}^{in}(m)$, thus vacuum input mode $\hat{a}_{L}^{in}(m)$ should be considered. 
  For the same reason, vacuum input mode $\hat{a}_{R}^{in}(m+N+1)$ should also be added. 
  Such a multiple input BS process can be constructed from a single input case (path $m$), then adding the next input mode accordingly. Each time we add a new input mode, a new parallel BS process is established, meaning a new pair of input modes $\hat{a}_{L}^{in}(j)$ and $\hat{a}_{R}^{in}(j+1)$ will contribute to the new output modes. 
 Adding them in this way,  the effective interaction Hamiltonian and corresponding time evolution operator for the $N$ parallel BS processes are
\begin{subequations}\label{H2}
\begin{align}
&\hat{H}_{eff}=\sum_{j=m}^{m+N}\hat{H}_j,\\
&\hat{S}(t)=e^{-\frac{i}{\hbar}\hat{H}_{eff}t}=\prod_{j=m}^{m+N}\hat{S}_j(t),
\end{align}
\end{subequations}
where $\hat{H}_j$ and $\hat{S}_j(t)=\exp(-i\hat{H}_jt/\hbar)$ is the effective interaction Hamiltonian and corresponding time evolution operator for the $j$-th parallel BS process.

Once we obtain the effective Hamiltonian of this parallel BS process, the transformation between input and output operators can be obtained sequentially. With the time evolution operator in Eq.~(\ref{H2}b),  $\hat{a}_{out}(j)=\hat{S}^\dag(t)\hat{a}_{in}(j)\hat{S}(t)$. For the general input 
$\hat{a}_{in}(j)=\alpha_j\hat{a}_{L}^{in}(j)+\beta_j\hat{a}_{R}^{in}(j)$, the output in the $j$ path is
  $\hat{a}_{out}(j)=\cos\Delta\cdot\hat{a}_{in}(j)+i\alpha_j\sin\Delta\cdot\hat{a}_{R}^{in}(j+1)+i\beta_j\sin\Delta\cdot\hat{a}_{L}^{in}(j-1)$. The full transformation can be written as
\begin{equation}\label{QBS5}
\tiny
\left[
\begin{array}{l}
\hat{a}_{R}^{out}(m+N+1)\\
~\\
\hat{a}_{out}(m+N)\\
~\\
\hat{a}_{out}(m+N-1)\\
~~~~~~\vdots\\
~~~~~~\vdots\\
\hat{a}_{out}(m+1)\\
~\\
\hat{a}_{L}^{out}(m)
\end{array}
\right]
=
\left[
\begin{array}{lllll}
\hat{J}_{0}&\hat{J}_{+}&~&~&\\
~&~&~&~&\\
\hat{J}_{_{-,m+N}}&\hat{J}_{0}&\hat{J}_{_{+,m+N}}&~&\\
~&~&~&~&\\
~&\hat{J}_{_{-,m+N-1}}&\hat{J}_{0}&\hat{J}_{_{+,m+N-1}}&\\
~&~~~~~~~\ddots&~~~~~~~\ddots&~~~~~\ddots&\\
~&~&\ddots&\ddots&\ddots\\
~&~&~~~~~\hat{J}_{_{-,m+1}}&~~~~~~\hat{J}_{0}&~~~~~~\hat{J}_{_{+,m+1}}\\
~&~&~&~&\\
~&~&~&~~~~~~~\hat{J}_{-}&~~~~~~\hat{J}_{0}
\end{array}
\right]
\left[
\begin{array}{l}
\hat{a}_{R}^{in}(m+N+1)\\
~\\
\hat{a}_{in}(m+N)\\
~\\
\hat{a}_{in}(m+N-1)\\
~~~~~~\vdots\\
~~~~~~\vdots\\
\hat{a}_{in}(m+1)\\
~\\
\hat{a}_{L}^{in}(m)
\end{array}
\right],
\end{equation}
where $\hat{J}_{+,l}=\beta_l\hat{J}_{+}$, $\hat{J}_{-,l}=\alpha_l\hat{J}_{-}$ with the index $l=m+1,m+2,\hdots,m+N$. Here, the column vectors contain $N+2$ annihilation operators, and the dimension of transformation matrix is $(N+2)\times(N+2)$. Equivalently, the transformation can be expressed in terms of LCP/RCP modes
\begin{equation}\label{QBS6}
\tiny
\left[
\begin{array}{l}
\hat{a}_{R}^{out}(m+N+1)\\
\hat{a}_{L}^{out}(m+N)\\
\hat{a}_{R}^{out}(m+N)\\
\hat{a}_{L}^{out}(m+N-1)\\
\hat{a}_{R}^{out}(m+N-1)\\
~~~~~~\vdots\\
\hat{a}_{L}^{out}(m+1)\\
\hat{a}_{R}^{out}(m+1)\\
\hat{a}_{L}^{out}(m)~~
\end{array}
\right]
=
\begin{bmatrix}
\cos\Delta&i\sin\Delta&~&~&~&~&~&~&~\\
i\sin\Delta&\cos\Delta&~&~&~&~&~&~&~\\
~&~&\cos\Delta&i\sin\Delta&~&~&~&~&~\\
~&~&i\sin\Delta&\cos\Delta&~&~&~&~&~\\
&~&~~&~&\cos\Delta&i\sin\Delta&~&~&~\\
~&~&~&~&~&\ddots&~&~&~\\
~&~&~&~&~&i\sin\Delta&\cos\Delta&~&~\\
~&~&~&~&~&~&~&\cos\Delta&i\sin\Delta\\
~&~&~&~&~&~&~&i\sin\Delta&\cos\Delta
\end{bmatrix}
\left[
\begin{array}{l}
\hat{a}_{R}^{in}(m+N+1)\\
\hat{a}_{L}^{in}(m+N)\\
\hat{a}_{R}^{in}(m+N)\\
\hat{a}_{L}^{in}(m+N-1)\\
\hat{a}_{R}^{in}(m+N-1)\\
~~~~~~\vdots\\
\hat{a}_{L}^{in}(m+1)\\
\hat{a}_{R}^{in}(m+1)\\
\hat{a}_{L}^{in}(m)~~
\end{array}
\right],
\end{equation}
where the column vectors contain $2(N+1)$ annihilation operators, and the dimension of transformation matrix is $2(N+1)\times2(N+1)$. The quantum state transformation can also be obtained directly with the time evolution operator (\ref{H2}b) by $|\psi\rangle_{out}=\hat{S}(t)|\psi\rangle_{in}$~\cite{QBS-book}.

Whether the one-to-multiple or two-to-two BS process, all previous studies treated the metasurface as a single BS process to realize the conversion or projection of the polarization state~\cite{MQP4,MQBS1,MQBS2} and to manipulate two-photon interference~\cite{MQM2,MQM3,MQBS1}.
But here, a gradient metasurface can be regarded as a series of cascading BS processes to act together, i.e., a parallel BS.
It will provide a powerful way to manipulate the quantum state even with multi-phonon case.  
In the next section, using the parallel BS on a metasurface, we will show the preparation and fusion of quantum entanglement by manipulating multiphoton quantum states. 


\section{Quantum state manipulation with parallel BS metasurface}

Before the discussion, we briefly introduce the notations we used to describe quantum states. 
A single photon state occupying path-$j$ mode with a certain polarization is denoted by Dirac notation $|A\rangle_j=\hat{a}_A^\dag(j)|0\rangle$ ($A=L,R,H,V$), where  $H$ and $V$ represent the horizontal and vertical linear polarization, respectively.
Here $\hat{a}_{H}(j)=[\hat{a}_{L}(j)+\hat{a}_{R}(j)]/\sqrt{2}$ and $\hat{a}_{V}(j)=i[\hat{a}_{L}(j)-\hat{a}_{R}(j)]/\sqrt{2}$.
If in the $j$ path,  two photons with LCP and RCP are occupied, it can be written as  $|LR\rangle_j=\hat{a}_L^\dag(j)\hat{a}_R^\dag(j)|0\rangle$, while there is a LCP photon in the $j$ path and RCP the $k$ path, $|L\rangle_j|R\rangle_{k}=\hat{a}_L^\dag(k)\hat{a}_R^\dag(k)|0\rangle$. With above relations, the operator transformation for inputting linearly polarized modes  become
\begin{subequations}\label{QBSHV}
\small
\begin{align}
\hat{a}_{R}^{out}(j)&=\frac{\cos\Delta}{\sqrt{2}}\cdot\hat{a}_{H}^{in}(j)+\frac{i\cos\Delta}{\sqrt{2}}\cdot\hat{a}_{V}^{in}(j)
+\frac{i\sin\Delta}{\sqrt{2}}\cdot\hat{a}_{H}^{in}(j-1)+\frac{\sin\Delta}{\sqrt{2}}\cdot\hat{a}_{V}^{in}(j-1)\\
\hat{a}_{L}^{out}(j)&=\frac{\cos\Delta}{\sqrt{2}}\cdot\hat{a}_{H}^{in}(j)-\frac{i\cos\Delta}{\sqrt{2}}\cdot\hat{a}_{V}^{in}(j)
+\frac{i\sin\Delta}{\sqrt{2}}\cdot\hat{a}_{H}^{in}(j+1)-\frac{\sin\Delta}{\sqrt{2}}\cdot\hat{a}_{V}^{in}(j+1)
\end{align}
\end{subequations}
With Eqs.~(\ref{QBS1}) and (\ref{QBSHV}) and their inverse transformation, one can easily obtain that $|\psi_{out}\rangle=\hat{S}(t)|\psi_{in}\rangle$. In the following, we will use above transformations to demonstrate the preparation and fusion of entangled states under a parallel BS of gradient metasurface.

{\color{red}
}

\subsection{Entangled state preparation}
\begin{figure}[htbp]
  \centering
  \includegraphics[width=0.50\textwidth]{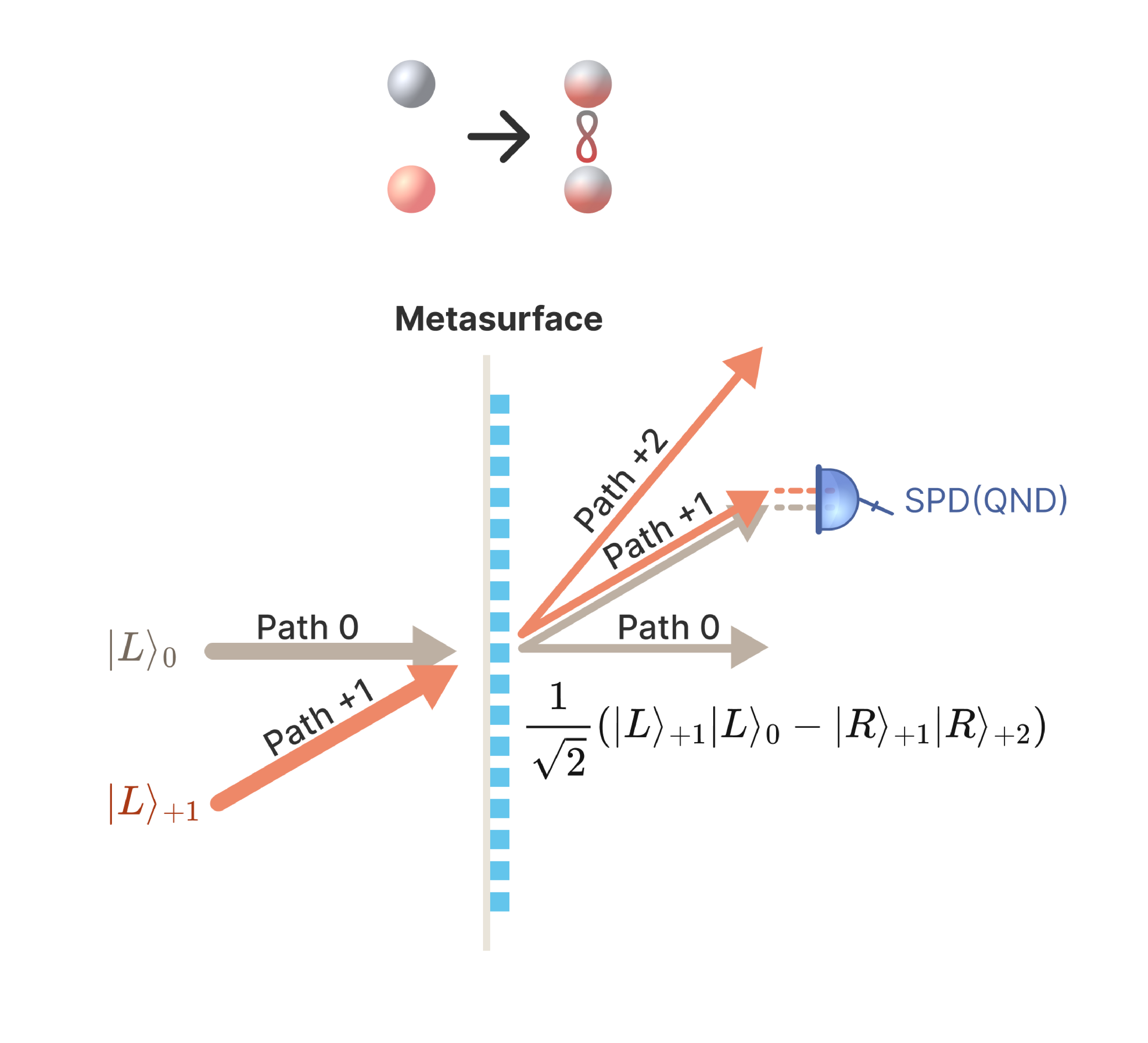}\\
  \caption{The schematic of two-photon entangled state preparation with parallel BS of metasurface. 
  When two LCP photons along paths 0 and +1 are incident to the metasurface, through a post-selection, the entangled state $\frac{1}{\sqrt{2}}(L\rangle_0|L\rangle_{+1}-|R\rangle_{+1}|R\rangle_{+2})$ is prepared. Here, SPD is an abbreviation of single photon detector.
}\label{fig8}
\end{figure}

We first discuss the preparation of two-photon entangled states. 
As shown in Fig.~\ref{fig8}, two LCP photons along paths 0 and +1 are sent into the metasurface, i.e., $|\psi_{in}\rangle=|L\rangle_0|L\rangle_{+1}$.  
In this case, at most four paths involve the transformation, so $H_{eff}=-g\hbar\sum_{j=-1}^{2}[\hat{a}_{L}^\dag(j)\hat{a}_{R}^\dag(j+1)+\hat{a}_{R}^\dag(j+1)\hat{a}_{L}^\dag(j)]$. With the relation $|\psi_{out}\rangle=\hat{S}(t)|\psi_{in}\rangle$, $|\psi_{out}\rangle$ reads
\begin{eqnarray}\label{psi_LL}
|\psi_{out}\rangle&&=\cos^2\Delta|L\rangle_0|L\rangle_{+1}-\sin^2\Delta|R\rangle_{+2}|R\rangle_{+1}
+i\sin\Delta\cos\Delta(|LR\rangle_{+1}+|L\rangle_0|R\rangle_{+2})
\end{eqnarray}
One can see that the first two terms in Eq.~(\ref{psi_LL}), i.e., $|\psi_{out}\rangle_{post}=\cos^2\Delta|L\rangle_0|L\rangle_{+1}-\sin^2\Delta|R\rangle_{+2}|R\rangle_{+1}$, have a form of entanglement.
The entangled component $|\psi_{out}\rangle_{post}$ can be post-selected if one and only one photon is detected in path +1, which can be done via quantum non-demolition measurement (QND) of photon number~\cite{QND-book,QND-article}. 
The success probability of post-selection is maximized (50\%) when two superposition coefficients have the same amplitude $|\cos^2\Delta|=|-\sin^2\Delta|=1/2$ , such as $\Delta=\pi/4$. What's more, leveraging the principle of path identity~\cite{Path-identity}, one can convert the state $|\psi_{out}\rangle_{post}$ into a maximally entangled state by combining the path (0, +2) into a single path.
The post-selected entangled state is a consequence of quantum interference, which is contributed by two simultaneous BS processes in the metasurface. 
Note that such a quantum interference effect is different from that arises in bulky beam splitters~\cite{QBS-R2,QBS-R3} or waveguide directional couplers~\cite{QBS-N2}.

\begin{figure*}[htbp]
  \centering
  \includegraphics[width=0.95\textwidth]{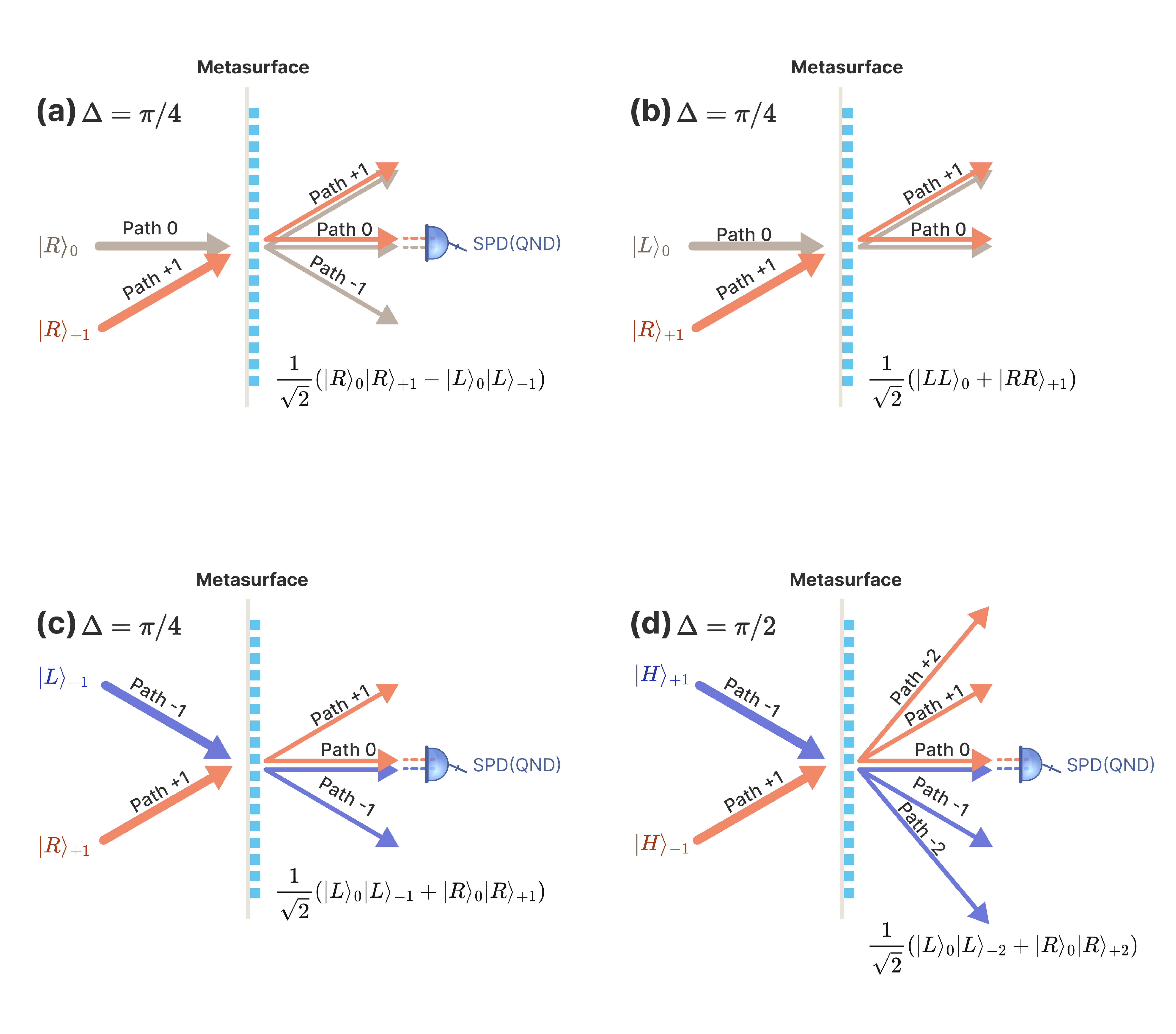}\\
  \caption{The schematic of 4 kinds of two-photon entangled state preparation with parallel BS of gradient metasurface.  (a) $|\psi_{in}\rangle=|R\rangle_{0}|R\rangle_{+1}$, $|\psi_{out}\rangle_{post}=(|R\rangle_{0}|R\rangle_{+1}-|L\rangle_{0}|L\rangle_{-1})/\sqrt{2}$; 
  (b) $|\psi_{in}\rangle=|L\rangle_{0}|R\rangle_{+1}$, $|\psi_{out}\rangle_{post}=(|LL\rangle_{0}+|RR\rangle_{+1})/\sqrt{2}$; 
  (c) $|\psi_{in}\rangle=|L\rangle_{-1}|R\rangle_{+1}$, $|\psi_{out}\rangle_{post}=(|L\rangle_{0}|L\rangle_{-1}+|R\rangle_{0}|R\rangle_{+1})/\sqrt{2}$; 
  (d) $|\psi_{in}\rangle=|H\rangle_{-1}|H\rangle_{+1}$, $|\psi_{out}\rangle_{post}=(|L\rangle_{0}|L\rangle_{-2}+|R\rangle_{0}|R\rangle_{+2})/\sqrt{2}$. The splitting parameters are $\Delta=\pi/4$ for (a), (b), and (c), and $\Delta=\pi/2$ for (d). 
  The post-selection  with success probability 50\% is done by single photon detection in path 0 for  (a), (c), and (d), while for (b) the post-selection is not required.
}\label{fig9}
\end{figure*}

Through choosing different input states, one can fabricate several types of two-photon entanglement states.
 In the frame of Fig. 8, if only at most two photons exist, the general form of effective Hamiltonian can be written as $H_{eff}=-g\hbar\sum_{j=-2}^{2}[\hat{a}_{L}^\dag(j)\hat{a}_{R}^\dag(j+1)+\hat{a}_{R}^\dag(j+1)\hat{a}_{L}^\dag(j)]$.
The splitting parameters are $\Delta=\pi/4$ for (a), (b), and (c), and $\Delta=\pi/2$ for (d). 
In Fig. 8(a), if $|\psi_{in}\rangle=|R\rangle_{0}|R\rangle_{+1}$, through the transformation $|\psi_{out}\rangle=\hat{S}(t)|\psi_{in}\rangle$,  $|\psi_{out}\rangle_{post}=(|R\rangle_{0}|R\rangle_{+1}-|L\rangle_{0}|L\rangle_{-1})/\sqrt{2}$, which is achieved by single photon QND in path 0 with success probabilities 50\%.
While in Fig. 8(b), $|\psi_{in}\rangle=|L\rangle_{0}|R\rangle_{+1}$, then  $|\psi_{out}\rangle_{post}=(|LL\rangle_{0}+|RR\rangle_{+1})/\sqrt{2}$ without any need of the post-selection process, i.e., post-selection free.
In Figs. 8(a) and 8(b), two BSs together are involved in the fabrication process of entanglement.
Then for Fig. 8(c), if $|\psi_{in}\rangle=|L\rangle_{-1}|R\rangle_{+1}$, $|\psi_{out}\rangle_{post}=(|L\rangle_{0}|L\rangle_{-1}+|R\rangle_{0}|R\rangle_{+1})/\sqrt{2}$; and for Fig. 8(d), $|\psi_{in}\rangle=|H\rangle_{-1}|H\rangle_{+1}$, then $|\psi_{out}\rangle_{post}=(|L\rangle_{0}|L\rangle_{-2}+|R\rangle_{0}|R\rangle_{+2})/\sqrt{2}$.
Both two processes in Figs. 8(c) and 8(d) need three BSs and their success probability through the post-selection is 50\%.
Thus, combining the unique parallel BS property of metasurface with a suitable post-selection together enables various two-photon entangled state preparations.

By adding the third photon into an adjacent path, we extend the two-photon case in Fig.~8(a) into three-photon case in Fig.~9(a).
When the input state is $|\psi_{in}\rangle=|R\rangle_{-1}|R\rangle_{0}|R\rangle_{+1}$, with $H_{eff}=-g\hbar\sum_{j=-2}^{2}[\hat{a}_{L}^\dag(j)\hat{a}_{R}^\dag(j+1)+\hat{a}_{R}^\dag(j+1)\hat{a}_{L}^\dag(j)]$ and $|\psi_{out}\rangle=\hat{S}(t)|\psi_{in}\rangle$, $|\psi_{out}\rangle$ reads
\begin{widetext}
\begin{eqnarray}\label{psi_RRR}
|\psi_{out}\rangle&&=\cos^3\Delta|R\rangle_{-1}|R\rangle_0|R\rangle_{+1}-i\sin^3\Delta|L\rangle_{-1}|L\rangle_{0}|L\rangle_{-2}\nonumber\\
&&+i\sin\Delta\cos^2\Delta(|LR\rangle_0|R\rangle_{-1}+|LR\rangle_{-1}|R\rangle_{+1}+|L\rangle_{-2}|R\rangle_0|R\rangle_{+1})\nonumber\\
&&-\cos\Delta\sin^2\Delta(|LR\rangle_0|L\rangle_{-2}+|LR\rangle_{-1}|L\rangle_{0}+|L\rangle_{-2}|L\rangle_{-1}|R\rangle_{+1})
\end{eqnarray}
\end{widetext}
One can see that the first two components in Eq.~(\ref{psi_RRR}), $|\psi_{out}\rangle_{post}=\cos^3\Delta|R\rangle_{-1}|R\rangle_0|R\rangle_{+1}-i\sin^3\Delta|L\rangle_{-1}|L\rangle_{0}|L\rangle_{-2}$, have polarization-polarization entanglement and polarization-path entanglement forms simultaneously.
$|\psi_{out}\rangle_{post}$ can be obtained if one and only one photon is post-selected in both paths 0 and  +1.
 The success properbality to prepare $|\psi_{out}\rangle_{post}$ is maximized to 25\% when two superposition coefficients have the same amplitude $|\cos^3\Delta|=|-i\sin^3\Delta|$ as $\Delta=\pi/4$, corresponding to a split ration of 50:50. 
 Similar to the two-photon case, the post-selected entanglement comes from three simultaneous BS processes in the metasurface. Also, $|\psi_{out}\rangle_{post}$ can be converted into a GHZ-type maximum entangled state after combining paths (-2,+1) due to path identity~\cite{Path-identity}.
\begin{figure*}[htbp]
  \centering
  \includegraphics[width=0.95\textwidth]{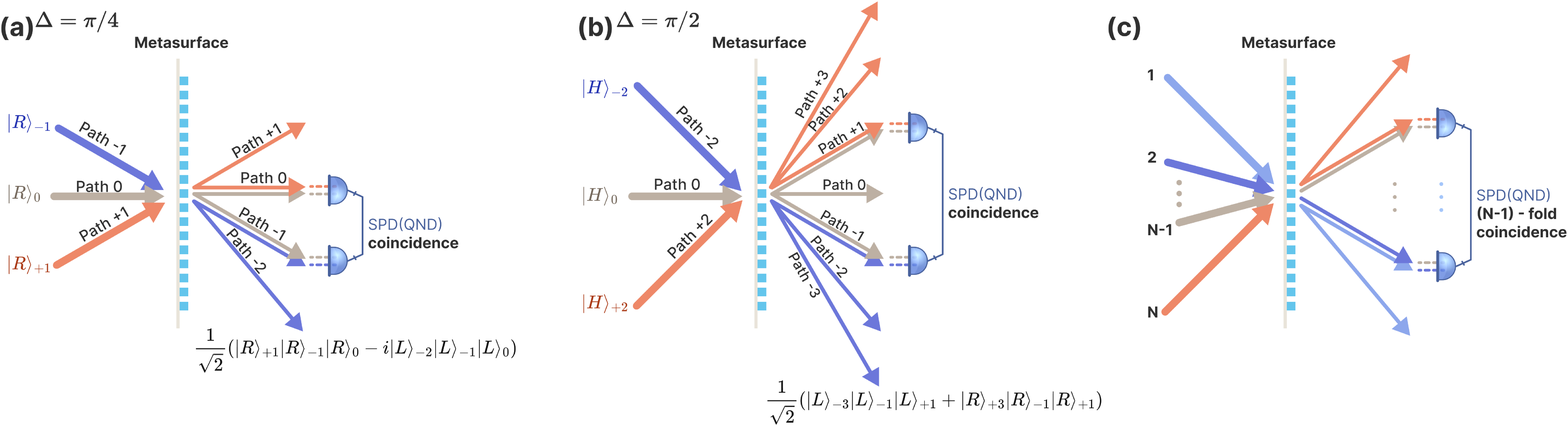}\\
  \caption{The schematic of three-photon entangled state preparations with parallel BS of gradient metasurface. 
  (a)Three-photon entangled state preparation with input state $|\psi_{in}\rangle=|R\rangle_{-1}|R\rangle_{0}|R\rangle_{+1}$, which needs single photon detection coincidence in paths 0 and  -1 for post-selection. (b) Three-photon entangled state preparation with input state $|\psi_{in}\rangle=|H\rangle_{-2}|H\rangle_{0}|H\rangle_{+2}$, which needs single photon detection coincidence in paths +1 and -1 for post-selection.
  The splitting parameters for (a) and (b) are $\Delta=\pi/4$ and $\Delta=\pi/2$, respectively.
   (c) The extended scheme of Fig.~\ref{fig9}(a) or Fig.~\ref{fig9}(d) for the preparation of $N$-photon entangled state, which needs ($N-1$)-fold single photon detection coincidence for post-selection.  
}\label{fig10}
\end{figure*}

Just like in the two-photon entanglement process, different setups of parallel BS can fabricate different types of three-photon entangled states. As shown in Fig.~\ref{fig10}(b), which is an extension of Fig.~\ref{fig9}(d),  three linearly polarized photons with path indexes differing by 2 are the input: $|\psi_{in}\rangle=|H\rangle_{-2}|H\rangle_{0}|H\rangle_{+2}$.
Then the entangled state is obtained under $\Delta=\pi/2$ with a success probability 1/4, which reads
$|\psi_{out}\rangle_{post}=(|L\rangle_{-3}|L\rangle_{-1}|L\rangle_{+1}+|R\rangle_{+3}|R\rangle_{-1}|R\rangle_{+1})/2\sqrt{2}$. 
It is noted that the three photons are entangled in both polarization-polarization and path-polarization forms because the path and polarization are locked for the first photon, so the dimension of entanglement is greatly improved.
This post-selected state can also be transformed into a GHZ-type entangled state by combining paths (-3,+3). Superior to previous three-photon GHZ preparation, which requires both entangled photon pairs and multiple beam splitters~\cite{QBS-R3}, this parallel BS method enabled by a single metasurface is entanglement resource-free and highly integrable.


Using the principle of parallel BS in gradient metasurface, $N$-photon entanglement can also be prepared. 
Here, we only give two types of entanglement. As shown in Fig.~\ref{fig10}(c), the generation of entanglement is accomplished by simultaneously detecting one photon in each of $(N-1)$ overlapped output paths after inputting $N$ single photons. 
For the first type, the input state is $|\psi_{in}\rangle=\prod_{j=1}^{N}|R\rangle_{m+j}=|R\rangle_{m+1}|R\rangle_{m+2}\cdot\cdot\cdot|R\rangle_{m+N}$. 
Then, the metasurface transforms the input state into $|\psi_{out}\rangle=\hat{S}(t)|\psi_{in}\rangle$, which reads
\begin{eqnarray}\label{psi_NR}
|\psi_{out}\rangle&&=\cos^N\Delta\prod_{j=1}^{N}|R\rangle_{m+j}+i^N\sin^N\Delta\prod_{j=0}^{N-1}|L\rangle_{m+j}
+|\psi_{discard}\rangle,
\end{eqnarray}
where $|\psi_{discard}\rangle$ is the discarded component after the post-selection. The first two terms in Eq.~(\ref{psi_NR}) form a $N$-photon entangled state, which can be post-selected by single photon QND of paths $(m+1,m+2,..., m+N-1)$ with a successful probability $\cos^{2N}\Delta+\sin^{2N}\Delta$. When the two coefficients have the same amplitude, i.e., $\Delta=\pi/4$, the success probability is maximized to $1/2^{N-1}$. Combining paths ($m,m+N$), the state becomes a GHZ-type polarization entangled state.

For the second type of $N$-photon entanglement, if the input state is $|\psi_{in}\rangle=\prod_{j=1}^{N}|H\rangle_{m+2j}=|H\rangle_{m+2}|H\rangle_{m+4}\cdot\cdot\cdot|H\rangle_{m+2N}$, then $|\psi_{out}\rangle$ reads
\begin{eqnarray}\label{psi_NH}
|\psi_{out}\rangle&&=\frac{i^N}{(\sqrt{2})^N}\prod_{j=1}^{N}|L\rangle_{m+2j-1}+\frac{i^N}{(\sqrt{2})^N}\prod_{j=1}^{N}|R\rangle_{m+2j+1}
+|\psi_{discard}\rangle
\end{eqnarray}
Here, we use the BS parameter $\Delta=\pi/2$ to maximize the post-selection probability. The first two terms in Eq.~(\ref{psi_NH}) also superpose a $N$-photon entangled state, which can be post-selected by single photon QND of paths $(m+3,m+5,…m+2N-1$) with a success probability $1/2^{N-1}$. 
In both cases, the metasurface provides $N$ parallel BS processes with every two adjacent BSs sharing a common output path.
When  $N=2,3$, the results in Eqs.~(\ref{psi_NR}) and (\ref{psi_NH}) are back to our previous discussion.

Recently, multiphoton entangled state generation has made great progress with compact metalens~\cite{MQP1} and resonant metasurface~\cite{MQP3}. Nevertheless, a nonlinear optical response is needed, which needs a trade-off between efficiency and integration. 
With a parallel BS principle, such a limitation is eliminated to some extent.
Thus,  gradient metasurface with parallel BS provides a powerful way for entangled state preparation.

\subsection{Entanglement fusion}
\begin{figure*}[htbp]
  \centering
  \includegraphics[width=0.95\textwidth]{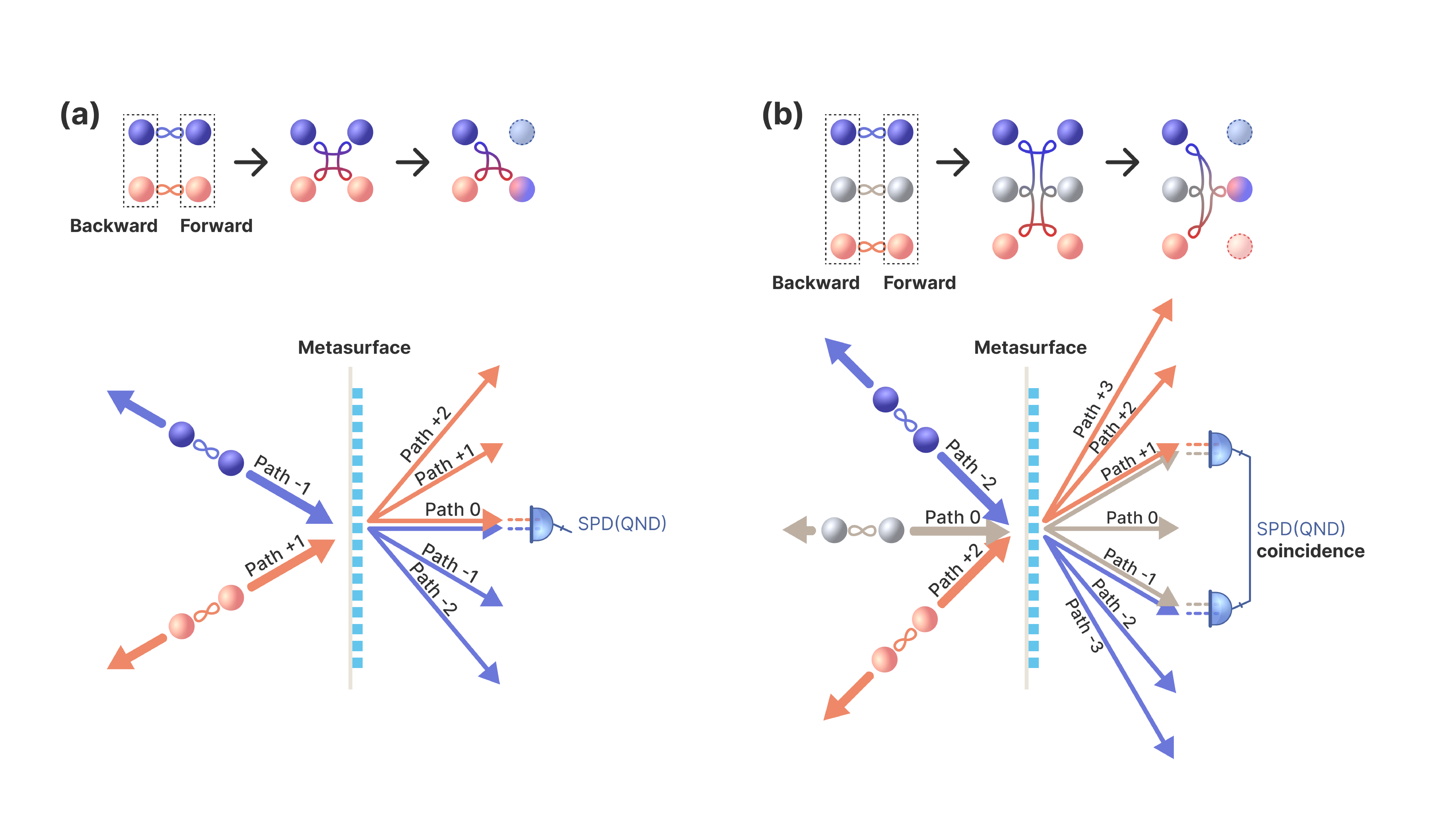}\\
  \caption{The schematic of entanglement fusion and conversion with parallel BS of a metasurface. (a) Fusion and conversion of two pairs of Bell states. (b) Fusion and conversion of three pairs of Bell states. 
  Here, the BS parameter of metasurface is $\Delta=\pi/2$.
}\label{fig11}
\end{figure*}
We have shown that the parallel BS metasurface can entangle the previously unentangled photons in a probable way, i.e., entanglement projection. But its successful probability will decrease quickly as the number of entangled photons increases, which limits the scale of entangled states. 
The entanglement fusion, which can merge several groups of independent entangled states into a large-scale entangled state~\cite{QBS-F1}, is helpful to perform universal quantum computing~\cite{QBS-F2}.
The unique manner of quantum interference in parallel BS  makes it possible to realize entanglement fusion and conversion, which will be demonstrated subsequently.

Firstly, consider the fusion and conversion of two pairs of Bell states. 
The key is to setup the connection between these two entangled pairs.
To this end, one of photons from one pair of the same path is passing through the metasurface to interfere with one of photons from another pair in another path, while another two photons are not involved in the interference process.
  Here, we use the polarization entangled state as the resources for fusion, which has the form $|\Psi^+(m)\rangle=(|H\rangle_{m}|V\rangle_{B,m}+|V\rangle_{m}|H\rangle_{B,m})/\sqrt{2}=(|L\rangle_{m}|L\rangle_{B,m}-|R\rangle_{m}|R\rangle_{B,m})/\sqrt{2}i$, where the subscript $B$ means backward, denoting that the photon is not through the metasurface, and $m$ marks the occupied path mode.
As shown in Fig.~\ref{fig11}(a), the input state is $|\psi_{in}\rangle=|\Psi^+(-1)\rangle|\Psi^+(+1)\rangle$. With the transformation $|\psi_{out}\rangle=\hat{S}(t)|\psi_{in}\rangle$, the output state reads
\begin{eqnarray}\label{Fusion_2LR}
|\psi_{out}\rangle&&=\frac{1}{2}|R\rangle_{+2}|R\rangle_{0}|L\rangle_{B,-1}|L\rangle_{B,+1}+\frac{1}{2}|L\rangle_{-2}|L\rangle_{0}|R\rangle_{B,-1}|R\rangle_{B,+1}+|\psi_{discard}\rangle.
\end{eqnarray}
Here, $\Delta=\pi/2$, and $|\psi_{discard}\rangle$ can be discarded through post-selection.
The first two terms in Eq.~(\ref{Fusion_2LR}) have the form of four-photon entangled state, which means the initial two pairs of Bell states can be fused into a genuine four-photon entangled state with a success probability $1/2$. Note that the post-selected state can be changed into a GHZ-type polarization entangled state if one combines paths (+2,-2).
The entanglement fusion obtained by metasurface originates from two parallel BS processes (red and dark blue arrows in Fig.~\ref{fig11}(a)). 

The fusion process through metasurface is superior to those reported in Ref.~\cite{QBS-F1, QBS-F2}, that is, it is not necessary to pay the photon number decreasing as the cost. The reason is that the QND for post-selection in our scheme is a non-destructive detection that preserves polarization information. 
 If expanding the detected photon state on a linear polarized basis, i.e., 
\begin{eqnarray}\label{Fusion_2HV}
|\psi_{out}\rangle&&=\frac{1}{2\sqrt{2}}|H\rangle_{0} |R\rangle_{+2}|L\rangle_{B,-1}|L\rangle_{B,+1}+\frac{1}{2\sqrt{2}}|H\rangle_{0}|L\rangle_{-2}|R\rangle_{B,-1}|R\rangle_{B,+1}\nonumber\\
&&-\frac{i}{2\sqrt{2}}|V\rangle_{0} |R\rangle_{+2}|L\rangle_{B,-1}|L\rangle_{B,+1}+\frac{i}{2\sqrt{2}}|V\rangle_{0}|L\rangle_{-2}|R\rangle_{B,-1}|R\rangle_{B,+1}\nonumber\\
&&+|\psi_{discard}\rangle,
\end{eqnarray}
the destructive detection is also available for our scheme.
Now, if one H (or V)-polarized photon is detected in path 0, then we obtain a three-photon entangled state with success probability 1/2.
In this fusion process, only two sets of BS are involved, so the power of parallel BS is not fully displayed. 

Then, the fusion and conversion of three pairs of Bell states are investigated. 
As shown in Fig.~\ref{fig11}(b),  only one of photons from each pair is passing through the metasurface, while another photon of each pair is propagating backward.
If the input state is $|\psi_{in}\rangle=|\Psi^+(-2)\rangle|\Psi^+(0)\rangle|\Psi^+(+2)\rangle$, as the splitting parameter $\Delta=\pi/2$, the quantum state becomes 
\begin{eqnarray}\label{Fusion_3LR}
|\psi_{out}\rangle&&=\frac{|R\rangle_{+3}|R\rangle_{-1}|R\rangle_{+1}|L\rangle_{B,-2}|L\rangle_{B,0}|L\rangle_{B,+2}}{2\sqrt{2}}\nonumber\\
&&-\frac{|L\rangle_{-3}|L\rangle_{-1}|L\rangle_{+1}|R\rangle_{B,-2}|R\rangle_{B,0}|R\rangle_{B,+2}}{2\sqrt{2}}+|\psi_{discard}\rangle.
\end{eqnarray}
The first two terms in Eq.~(\ref{Fusion_3LR}) can be post-selected by a coincident click of QND single photon detection in the paths $-1$ and $+1$. They can form a six-photon entangled state with a success probability of $1/4$. The fusion of three Bell pairs is achieved through a single gradient metasurface involving three parallel BSs.

By replacing the QND with destructive polarization detection, the fused six-photon entangled state will be converted into a four-photon entangled state. To see this, we change the circular polarization basics into linear ones, Eq.~(\ref{Fusion_3LR}) becomes
\begin{eqnarray}\label{Fusion_3HV}
|\psi_{out}\rangle&&=|H\rangle_{-1}|H\rangle_{+1}\frac{|R\rangle_{+3}|L\rangle_{B,-2}|L\rangle_{B,0}|L\rangle_{B,+2}-|L\rangle_{-3}|R\rangle_{B,-2}|R\rangle_{B,0}|R\rangle_{B,+2}}{4\sqrt{2}}\nonumber\\
&&-i|H\rangle_{-1}|V\rangle_{+1}\frac{|R\rangle_{+3}|L\rangle_{B,-2}|L\rangle_{B,0}|L\rangle_{B,+2}+|L\rangle_{-3}|R\rangle_{B,-2}|R\rangle_{B,0}|R\rangle_{B,+2}}{4\sqrt{2}}\nonumber\\
&&-i|V\rangle_{-1}|H\rangle_{+1}\frac{|R\rangle_{+3}|L\rangle_{B,-2}|L\rangle_{B,0}|L\rangle_{B,+2}+|L\rangle_{-3}|R\rangle_{B,-2}|R\rangle_{B,0}|R\rangle_{B,+2}}{4\sqrt{2}}\nonumber\\
&&-|V\rangle_{-1}|V\rangle_{+1}\frac{|R\rangle_{+3}|L\rangle_{B,-2}|L\rangle_{B,0}|L\rangle_{B,+2}-|L\rangle_{-3}|R\rangle_{B,-2}|R\rangle_{B,0}|R\rangle_{B,+2}}{4\sqrt{2}}\nonumber\\
&&+|\psi_{discard}\rangle.
\end{eqnarray}
If two H-polarized photons in the paths $-1$ and $+1$ are simultaneously detected, then the post-selected state becomes the first term in Eq.~(\ref{Fusion_3HV}), i.e., a four-photon entangled state. Also, the other three possible detection outcomes (HV, VH, VV) will yield a similar four-photon entangled state. 
Thus, the fusion of three Bell pairs can be completed in a single metasurface.

Above fusion and conversion scheme can be extended to a more general case with $N$ pairs of Bell states. 
The input state reads $|\psi_{in}\rangle=\prod_{j=1}^{N}|\Psi^+(m+2j)\rangle$. 
When half of photons pass through the metasurface, the quantum state becomes 
\begin{eqnarray}\label{Fusion_NLR}
|\psi_{out}\rangle&&=(\frac{1}{\sqrt{2}})^N\prod_{j=1}^{N}|R\rangle_{m+2j+1}|L\rangle_{B,m+2j}+(-\frac{1}{\sqrt{2}})^N\prod_{j=1}^{N}|L\rangle_{m+2j-1}|L\rangle_{B,m+2j}\nonumber\\
&&+|\psi_{discard}\rangle.
\end{eqnarray}
The first two terms in Eq.~(\ref{Fusion_NLR}) can be preserved, which constitutes a $2N$-photon entangled state. The post-selection condition is the same as the situation for $N$-photon entangled state generation, with a success probability $1/2^{N-1}$. Here $N$ BSs take part in the fusion process.
Thus, the parallel BS metasurface provides a compact platform for multipartite entanglement fusion and state conversion.
 
 The principle of above entanglement fusion is not limited to the Bell states. In fact, it can be extended to any entanglement states. So, through a gradient metasurface, it is possible to construct a large-scale cluster state by fusing several smaller cluster states, which is critical for measurement-based quantum computing~\cite{QBS-F1,QBS-F2}. 
 Also, the capability to establish the entanglement among multiple parties makes it possible to distribute quantum entanglement  or build a multi-party quantum network~\cite{QBS-S3}. 
So metasurface provides a compact way  to realize above quantum functions, which can be used  in  integrated quantum optics and quantum information processing.

Taking advantage of the multi-degree-of-freedom control capabilities, BS metasurface has shown great potential for quantum photonics, such as quantum state reconstruction~\cite{MQBS1}, multichannel entanglement distribution~\cite{MQBS2}, high-dimensional entanglement generation~\cite{MQM3}, etc. 
However, all these works  are only using the single BS process of metasurface. 
Here, the finding of parallel BS enhances the capability of metasurface for quantum light manipulation. 
 Such a scheme improves the parallelization of quantum manipulation, facilitating the multifunctionality of metasurface-based quantum devices.

\section{Summary}
In this work, we have revealed  the mechanism of parallel BS of PB-phase metasurfaces and demonstrated its application in entanglement manipulation. 
The parallel BS metasurfcaes can be modeled as a series linked beam splitters, based on which, we have set up the consistent and intuitive transformation rules of metasurface for both classical and quantum light.
More importantly, the parallel BS enables a powerful quantum interference capability for quantum state manipulation. 
So we have demonstrated that a multi-photon entangled state can be prepared under multiple parallel BS processes. Furthermore, the fusion of entanglement among multi-parity has been realized with a single compact metasurfce. 
The principle of parallel BS through the metasurface provides up a versatile way to manipulate the quantum state at the micro/nano scale. 
So the results present here may open up an avenue for planar quantum integration with metasurfaces, promoting the parallelization and multifunctionality of on-chip quantum information processing.

\acknowledgments
\textit{Acknowledgments.} This work is supported by the National Natural Science Foundation of China under Grants No. 11974032, No. 12161141010 and No. T2325022, by the Innovation Program for Quantum Science and Technology under Grant No. 2021ZD0301500, and by the Key R$\&$D Program of Guangdong Province under Grant No. 2018B030329001.



\begin{thebibliography}{99}
\bibitem{MSR1}	N. Yu, P. Genevet, M. A. Kats, F. Aieta, J.- P. Tetienne, F. Capasso, and Z. Gaburro, {\it Light Propagation with Phase Discontinuities: Generalized Laws of Reflection and Refraction}, Science \textbf{334}, 333 (2011).

\bibitem{MSR2}	H.- T. Chen, A. J. Taylor, and N. Yu, {\it A Review of Metasurfaces: Physics and Applications}, Rep. Prog. Phys. \textbf{79}, 076401 (2016).

\bibitem{MSR3}	C.- W. Qiu, T. Zhang, G. Hu, and Y. Kivshar, {\it Quo Vadis, Metasurfaces?}, Nano Lett. \textbf{21}, 5461 (2021).

\bibitem{MSR4}	A. H. Dorrah and F. Capasso, {\it Tunable Structured Light with Flat Optics}, Science \textbf{376}, 367 (2022).

\bibitem{MSPO2}	S.-C. Jiang, X. Xiong, Y.-S. Hu, Y.-H. Hu, G.-B. Ma, R.-W. Peng, C. Sun, and M. Wang, {\it Controlling the Polarization State of Light with a Dispersion-Free Metastructure}, Phys. Rev. X \textbf{4}, 021026 (2014).

\bibitem{MSPO3}	Z. Wu, Y. Ra’di, and A. Grbic, {\it Tunable Metasurfaces: A Polarization Rotator Design}, Phys. Rev. X \textbf{9}, 011036 (2019).

\bibitem{MSPO}	N. A. Rubin, G. D'Aversa, P. Chevalier, Z. Shi, W. T. Chen, and F. Capasso, {\it Matrix Fourier Optics Enables a Compact Full-Stokes Polarization Camera}, Science \textbf{365}, 43 (2019).
\bibitem{MSNO}	G. Li, S. Zhang, and T. Zentgraf, {\it Nonlinear Photonic Metasurfaces}, Nat. Rev. Mater. \textbf{2}, 17010 (2017).

\bibitem{MSSL1}	A. C. Overvig, S. A. Mann, and A. Al\`{u}, {\it Thermal Metasurfaces: Complete Emission Control by Combining Local and Nonlocal Light-Matter Interactions}, Phys. Rev. X \textbf{11}, 021050 (2021).

\bibitem{MSSL2}	H. Ahmed, H. Kim, Y. Zhang, Y. Intaravanne, J. Jang, J. Rho, S. Chen, and X. Chen, {\it Optical Metasurfaces for Generating and Manipulating Optical Vortex Beams}, Nanophotonics \textbf{11}, 941 (2022).

\bibitem{MSH1}	S. M. Kamali, E. Arbabi, A. Arbabi, Y. Horie, M. Faraji-Dana, and A. Faraon, {\it Angle-Multiplexed Metasurfaces: Encoding Independent Wavefronts in a Single Metasurface under Different Illumination Angles}, Phys. Rev. X \textbf{7}, 041056 (2017).

\bibitem{MSH2}	L. Huang, S. Zhang, and T. Zentgraf, {\it Metasurface Holography: From Fundamentals to Applications}, Nanophotonics \textbf{7}, 1169 (2018).

\bibitem{Lens2}	F. Monticone, C. A. Valagiannopoulos, and A. Al\`{u}, {\it Parity-Time Symmetric Nonlocal Metasurfaces: All-Angle Negative Refraction and Volumetric Imaging}, Phys. Rev. X \textbf{6}, 041018 (2016).

\bibitem{Lens}	A. Arbabi and A. Faraon, {\it Advances in Optical Metalenses}, Nat. Photonics \textbf{17}, 16 (2023).

\bibitem{MQR}	A. S. Solntsev, G. S. Agarwal, and Y. S. Kivshar, {\it Metasurfaces for Quantum Photonics}, Nat. Photonics \textbf{15}, 327 (2021).

\bibitem{MQP1}	L. Li, Z. Liu, X. Ren, S. Wang, V.- C. Su, M.- K. Chen, C. H. Chu, H. Y. Kuo, B. Liu, W. Zang, G. Guo, L. Zhang, Z. Wang, S. Zhu, and D. P. Tsai, {\it Metalens-Array-Based High-Dimensional and Multiphoton Quantum Source}, Science \textbf{368}, 1487 (2020).

\bibitem{MQP2}	W. J. M. Kort-Kamp, A. K. Azad, and D. A. R. Dalvit, {\it Space-Time Quantum Metasurfaces}, Phys. Rev. Lett. \textbf{127}, 043603 (2021).

\bibitem{MQP3}	T. Santiago-Cruz, S. D. Gennaro, O. Mitrofanov, S. Addamane, J. Reno, I. Brener, and M. V. Chekhova, {\it Resonant Metasurfaces for Generating Complex Quantum States}, Science \textbf{377}, 991 (2022).

\bibitem{MQP4}	Z.- X. Li, D. Zhu, P.- C. Lin, P.- C. Huo, H.- K. Xia, M.- Z. Liu, Y.- P. Ruan, J.- S. Tang, M. Cai, H.- D. Wu, C.- Y. Meng, H. Zhang, P. Chen, T. Xu, K.- Y. Xia, L.- J. Zhang, and Y.- Q. Lu, {\it High-Dimensional Entanglement Generation Based on a Pancharatnam-Berry Phase Metasurface}, Photonics Res. \textbf{10}, 2702 (2022).

\bibitem{MQM1}	T. Stav, A. Faerman, E. Maguid, D. Oren, V. Kleiner, E. Hasman, and M. Segev, {\it Quantum Entanglement of the Spin and Orbital Angular Momentum of Photons Using Metamaterials}, Science \textbf{361}, 1101 (2018).

\bibitem{MQM2}	P. Georgi, M. Massaro, K.- H. Luo, B. Sain, N. Montaut, H. Herrmann, T. Weiss, G. Li, C. Silberhorn, and T. Zentgraf, {\it Metasurface Interferometry toward Quantum Sensors}, Light-Sci. Appl. \textbf{8}, 70 (2019).

\bibitem{MQM3}	Q. Li, W. Bao, Z. Nie, Y. Xia, Y. Xue, Y. Wang, S. Yang, and X. Zhang, {\it A Non-Unitary Metasurface Enables Continuous Control of Quantum Photon-Photon Interactions from Bosonic to Fermionic}, Nat. Photonics \textbf{15}, 267 (2021).

\bibitem{MQM4}	D. Zhang, Y. Chen, S. Gong, W. Wu, W. Cai, M. Ren, X. Ren, S. Zhang, G. Guo, and J. Xu, {\it All-Optical Modulation of Quantum States by Nonlinear Metasurface}, Light-Sci. Appl. \textbf{11}, 58 (2022).

\bibitem{MQM5}	Z. Gao, Z. Su, Q. Song, P. Genevet, and K. E. Dorfman, {\it Metasurface for Complete Measurement of Polarization Bell State}, Nanophotonics \textbf{12}, 569 (2023).

\bibitem{MQM6}	H. Liang, H. Ahmed, W. Y. Tam, X. Chen, and J. Li, {\it Continuous Heralding Control of Vortex Beams Using Quantum Metasurface}, Commun. Phys. \textbf{6}, 140 (2023).

\bibitem{MQI1}	J. Zhou, S. Liu, H. Qian, Y. Li, H. Luo, S. Wen, Z. Zhou, G. Guo, B. Shi, and Z. Liu, {\it Metasurface Enabled Quantum Edge Detection}, Sci. Adv. \textbf{6}, eabc4385 (2020).

\bibitem{MQI2}	A. Vega, T. Pertsch, F. Setzpfandt, and A. A. Sukhorukov, {\it Metasurface-Assisted Quantum Ghost Discrimination of Polarization Objects}, Phys. Rev. Appl. \textbf{16}, 064032 (2021).

\bibitem{MQI3}	T. K. Yung, H. Liang, J. Xi, W. Y. Tam, and J. Li, {\it Jones-Matrix Imaging Based on Two-Photon Interference}, Nanophotonics \textbf{12}, 579 (2023).

\bibitem{MQG1}	P. K. Jha, X. Ni, C. Wu, Y. Wang, and X. Zhang, {\it Metasurface-Enabled Remote Quantum Interference}, Phys. Rev. Lett. \textbf{115}, 025501 (2015).

\bibitem{MQG2}	P. K. Jha, N. Shitrit, X. Ren, Y. Wang, and X. Zhang, {\it Spontaneous Exciton Valley Coherence in Transition Metal Dichalcogenide Monolayers Interfaced with an Anisotropic Metasurface}, Phys. Rev. Lett. \textbf{121}, 116102 (2018).

\bibitem{MQG3}	P. K. Jha, N. Shitrit, J. Kim, X. Ren, Y. Wang, and X. Zhang, {\it Metasurface-Mediated Quantum Entanglement}, ACS Photonics \textbf{5}, 971 (2018).

\bibitem{MSBS1}	A. Arbabi, Y. Horie, M. Bagheri, and A. Faraon, {\it Dielectric Metasurfaces for Complete Control of Phase and Polarization with Subwavelength Spatial Resolution and High Transmission}, Nat. Nanotechnol. \textbf{10}, 937 (2015).

\bibitem{MSBS2}	J. P. Balthasar Mueller, N. A. Rubin, R. C. Devlin, B. Groever, and F. Capasso, {\it Metasurface Polarization Optics: Independent Phase Control of Arbitrary Orthogonal States of Polarization}, Phys. Rev. Lett. \textbf{118}, 113901 (2017).

\bibitem{MSBS3}	M. Wei, Q. Xu, Q. Wang, X. Zhang, Y. Li, J. Gu, Z. Tian, X. Zhang, J. Han, and W. Zhang, {\it Broadband Non-Polarizing Terahertz Beam Splitters with Variable Split Ratio}, Appl. Phys. Lett. \textbf{111}, 071101 (2017).

\bibitem{MSBS4}	X. Zhang, R. Deng, F. Yang, C. Jiang, S. Xu, and M. Li, {\it Metasurface-Based Ultrathin Beam Splitter with Variable Split Angle and Power Distribution}, ACS Photonics \textbf{5}, 2997 (2018).

\bibitem{MSBS5}	V. S. Asadchy, A. D\'{i}az-Rubio, S. N. Tcvetkova, D.- H. Kwon, A. Elsakka, M. Albooyeh, and S. A. Tretyakov, {\it Flat Engineered Multichannel Reflectors}, Phys. Rev. X \textbf{7}, 031046 (2017).

\bibitem{MSBS6}	Y.- J. Gao, X. Xiong, Z. Wang, F. Chen, R.- W. Peng, and M. Wang, {\it Simultaneous Generation of Arbitrary Assembly of Polarization States with Geometrical-Scaling-Induced Phase Modulation}, Phys. Rev. X \textbf{10}, 031035 (2020).

\bibitem{MQBS1}	K. Wang, J. G. Titchener, S. S. Kruk, L. Xu, H.- P. Chung, M. Parry, I. I. Kravchenko, Y.- H. Chen, A. S. Solntsev, Y. S. Kivshar, D. N. Neshev, and A. A. Sukhorukov, {\it Quantum Metasurface for Multiphoton Interference and State Reconstruction}, Science \textbf{361}, 1104 (2018).

\bibitem{MQBS2}	Y.- J. Gao, Z. Wang, Y. Jiang, R.- W. Peng, Z.- Y. Wang, D.- X. Qi, R.- H. Fan, W.- J. Tang, and M. Wang, {\it Multichannel Distribution and Transformation of Entangled Photons with Dielectric Metasurfaces}, Phys. Rev. Lett. \textbf{129}, 023601 (2022).

\bibitem{QBS-R1}	C. K. Hong, Z. Y. Ou, and L. Mandel, {\it Measurement of Subpicosecond Time Intervals between 2 Photons by Interference}, Phys. Rev. Lett. \textbf{59}, 2044 (1987).

\bibitem{QBS-R2}	P. Kok, W. J. Munro, K. Nemoto, T. C. Ralph, J. P. Dowling, and G. J. Milburn, {\it Linear Optical Quantum Computing with Photonic Qubits}, Rev. Mod. Phys. \textbf{79}, 135 (2007).

\bibitem{QBS-R3}	J.- W. Pan, Z.- B. Chen, C.- Y. Lu, H. Weinfurter, A. Zeilinger, and M. \.{Z}ukowski, {\it Multiphoton Entanglement and Interferometry}, Rev. Mod. Phys. \textbf{84}, 777 (2012).

\bibitem{QBS-S1}	S. Bose, V. Vedral, and P. L. Knight, {\it Multiparticle Generalization of Entanglement Swapping}, Phys. Rev. A \textbf{57}, 822 (1998).

\bibitem{QBS-S2}	J.- W. Pan, D. Bouwmeester, H. Weinfurter, and A. Zeilinger, {\it Experimental Entanglement Swapping: Entangling Photons That Never Interacted}, Phys. Rev. Lett. \textbf{80}, 3891 (1998).

\bibitem{QBS-S3}	X. Su, C. Tian, X. Deng, Q. Li, C. Xie, and K. Peng, {\it Quantum Entanglement Swapping between Two Multipartite Entangled States}, Phys. Rev. Lett. \textbf{117}, 240503 (2016).

\bibitem{QBS-F1}	D. E. Browne and T. Rudolph, {\it Resource-Efficient Linear Optical Quantum Computation}, Phys. Rev. Lett. \textbf{95}, 010501 (2005).

\bibitem{QBS-F2}	S. Bartolucci, P. Birchall, H. Bomb\'{i}n, H. Cable, C. Dawson, M. Gimeno-Segovia, E. Johnston, K. Kieling, N. Nickerson, M. Pant, F. Pastawski, T. Rudolph, and C. Sparrow, {\it Fusion-Based Quantum Computation}, Nat. Commun. \textbf{14}, 912 (2023).

\bibitem{QBS-N1}	M. Reck, A. Zeilinger, H. J. Bernstein, and P. Bertani, {\it Experimental Realization of Any Discrete Unitary Operator}, Phys. Rev. Lett. \textbf{73}, 58 (1994).

\bibitem{QBS-N2}	J. Wang, F. Sciarrino, A. Laing, and M. G. Thompson, {\it Integrated Photonic Quantum Technologies}, Nat. Photonics \textbf{14}, 273 (2020).

\bibitem{QBS-N3}	N. Tischler, C. Rockstuhl, and K. S\l{}owik, {\it Quantum Optical Realization of Arbitrary Linear Transformations Allowing for Loss and Gain}, Phys. Rev. X \textbf{8}, 021017 (2018).

\bibitem{QBS-N4}	S. Barz, B. Daki\'{c}, Y. O. Lipp, F. Verstraete, J. D. Whitfield, and P. Walther, {\it Linear-Optical Generation of Eigenstates of the Two-Site $XY$ Model}, Phys. Rev. X \textbf{5}, 021010 (2015).

\bibitem{QBS-P1}	H.- S. Zhong, Y. Li, W. Li, L.- C. Peng, Z.- E. Su, Y. Hu, Y.- M. He, X. Ding, W. Zhang, H. Li, L. Zhang, Z. Wang, L. You, X.- L. Wang, X. Jiang, L. Li, Y.- A. Chen, N.- L. Liu, C.- Y. Lu, and J.- W. Pan, {\it 12-Photon Entanglement and Scalable Scattershot Boson Sampling with Optimal Entangled-Photon Pairs from Parametric Down-Conversion}, Phys. Rev. Lett. \textbf{121}, 250505 (2018).

\bibitem{QBS-P2}	W. Asavanant, Y. Shiozawa, S. Yokoyama, B. Charoensombutamon, H. Emura, R. N. Alexander, S. Takeda, J. Yoshikawa, N. C. Menicucci, H. Yonezawa, and A. Furusawa, {\it Generation of Time-Domain-Multiplexed Two-Dimensional Cluster State}, Science \textbf{366}, 373 (2019).

\bibitem{QBS-P3}	M. V. Larsen, X. Guo, C. R. Breum, J. S. Neergaard-Nielsen, and U. L. Andersen, {\it Deterministic Generation of a Two-Dimensional Cluster State}, Science \textbf{366}, 369 (2019).

\bibitem{BS-book}  E. Hecht, {\it Optics} (Pearson Education, Harlow, United Kingdom, 2013).

\bibitem{QBS-book} G. S. Agarwal, {\it Quantum optics} (Cambridge University Press, New York, 2012).

\bibitem{QND-book} K. Xia, in {\it Photon Counting-Fundamentals and Applications}, edited by B. Nikolay, and N. Anton (IntechOpen, Rijeka, 2018), Ch. 3.
\bibitem{QND-article}	J.-C. Besse, S. Gasparinetti, M. C. Collodo, T. Walter, P. Kurpiers, M. Pechal, C. Eichler, and A. Wallraff, {\it Single-Shot Quantum Nondemolition Detection of Individual Itinerant Microwave Photons}, Phys. Rev. X \textbf{8}, 021003 (2018).

\bibitem{Path-identity} L.-T. Feng, M. Zhang, D. Liu, Y.-J. Cheng, G.-P. Guo, D.-X. Dai, G.-C. Guo, M. Krenn, and X.-F. Ren, {\it On-chip Quantum Interference between the Origins of a Multi-photon State}, Optica \textbf{10}, 105–109 (2023).



\end{thebibliography}

\end{document}